\documentclass[journal,twoside]{IEEEtran}

\usepackage{color}
\usepackage{bbm}
\usepackage{amsmath,amsfonts,amssymb}
\usepackage{dsfont}
\usepackage{graphicx}
\usepackage[hang]{subfigure}
\usepackage{citesort}
\usepackage{mathtools}
\usepackage{bbm}
\usepackage{fixmath}
\usepackage{balance}

\usepackage{ntheorem}
\theoremstyle{plain}
\theoremheaderfont{\itshape}\theorembodyfont{\itshape}
\theoremseparator{.}
\newtheorem{theorem}{Theorem}
\newtheorem{corollary}{Corollary}

\newtheorem{lemma}{Lemma}

\theoremheaderfont{\itshape}\theorembodyfont{\upshape}
\theoremseparator{.}
\newtheorem{definition}{Definition}
\newtheorem{remark}{Remark}

\newcommand{\N}{\mathbb{N}}
\newcommand{\R}{\mathbb{R}}

\newcommand{\sJ}{\mathcal{J}}
\newcommand{\sK}{\mathcal{K}}

\newcommand{\sM}{\mathcal{M}}
\newcommand{\sP}{\mathcal{P}}

\newcommand{\sU}{\mathcal{U}}
\newcommand{\sV}{\mathcal{V}}

\newcommand{\sX}{\mathcal{X}}
\newcommand{\sY}{\mathcal{Y}}
\newcommand{\sZ}{\mathcal{Z}}

\newcommand{\Mn}{M_n}
\newcommand{\Kon}{K_{1,n}}
\newcommand{\Ktn}{K_{2,n}}
\newcommand{\Kin}{K_{i,n}}
\newcommand{\eo}{\bar{e}_{1,n}}
\newcommand{\et}{\bar{e}_{2,n}}
\newcommand{\ei}{\bar{e}_{i,n}}
\newcommand{\eps}{\epsilon_n}
\newcommand{\epsi}{\epsilon_{i,n}}
\newcommand{\epso}{\epsilon_{1,n}}
\newcommand{\epst}{\epsilon_{2,n}}
\newcommand{\del}{\delta_n}

\newcommand{\qed}{\hfill $\blacksquare$}

\begin{document}

\title{Secure Broadcasting Using Independent Secret Keys}
\author{Rafael F. Schaefer,~\IEEEmembership{Senior Member,~IEEE},
		Ashish Khisti,~\IEEEmembership{Member,~IEEE}, and
		H.~Vincent Poor,~\IEEEmembership{Fellow,~IEEE}
	\thanks{Manuscript received June 13, 2017; revised September 6, 2017; accepted October 9, 2017. Date of publication October 20, 2017; date of current version February 14, 2018.This work was supported in part by the U.S. National Science Foundation under Grants CNS-1702808 and ECCS-1647198. This paper was presented in part at the 48th Annual Conference on Information Sciences and Systems, Princeton, NJ, USA, March 2014 \cite{Schaefer14CommonMessageSecretKeys}, and the IEEE International Symposium on Information Theory, Hong Kong, June 2015 \cite{SchaeferKhistiPoor15SecretKeys}. The associate editor coordinating the review of this paper and approving it for publication was V. Stankovic. (\emph{Corresponding author: Rafael F. Schaefer.})}
	\thanks{R. F. Schaefer is with the Information Theory and Applications Chair, Technische Universit\"at Berlin, 10587 Berlin, Germany (e-mail: rafael.schaefer@tu-berlin.de).}
	\thanks{A. Khisti is with the Department of Electrical and Computer Engineering, University of Toronto, Toronto, ON M5S 3G4, Canada (e-mail: akhisti@ece.utoronto.ca).}	
	\thanks{H. V. Poor is with the Department of Electrical Engineering, Princeton University, Princeton, NJ 08544 USA (e-mail: poor@princeton.edu).}        
	\thanks{Color versions of one or more of the figures in this paper are available online at http://ieeexplore.ieee.org.}
	\thanks{Digital Object Identifier 10.1109/TCOMM.2017.2764892}
}

\IEEEspecialpapernotice{(Invited Paper)}
\IEEEoverridecommandlockouts
\maketitle

\markboth{IEEE Transactions on Communications, vol. 66, no. 2, February 2018}{Schaefer \MakeLowercase{\textit{et al.}}: Secure Broadcasting Using Independent Secret Keys}

\begin{abstract}
The problem of secure broadcasting with independent secret keys is studied. The particular scenario is analyzed in which a common message has to be broadcast to two legitimate receivers, while keeping an external eavesdropper ignorant of it. The transmitter shares independent secret keys of sufficiently high rates with both legitimate receivers, which can be used in different ways: they can be used as one-time pads to encrypt the common message, as fictitious messages for wiretap coding, or as a hybrid of these. In this paper, capacity results are established when the broadcast channels involving the three receivers are degraded. If both legitimate channels are degraded versions of the eavesdropper's channel, it is shown that the one-time pad approach is optimal for several cases, yielding corresponding capacity expressions. Alternatively, the wiretap coding approach is shown to be optimal if the eavesdropper's channel is degraded with respect to both legitimate channels, establishing capacity in this case as well. If the eavesdropper's channel is neither the strongest nor the weakest, an intricate scheme that carefully combines both concepts of one-time pad and wiretap coding with fictitious messages turns out to be capacity-achieving. Finally we also obtain some results for the general non-degraded broadcast channel.
\end{abstract}


\section{Introduction}
\label{sec:introduction}

Shannon was the first who studied in \cite{Shannon49CommunicationTheorySecrecySystems} the problem of secure communication from an information theoretic perspective. He considered a noiseless communication scenario, in which transmitter and receiver share a secret key that is unknown to an eavesdropper. Used as a \emph{one-time pad}, this secret key enables secure transmission of the confidential message. 

Subsequently, Wyner considered the noisy case in \cite{Wyner75WiretapChannel}, where he introduced the so-called \emph{wiretap channel}. He extended the problem studied by Shannon insofar that the legitimate receiver and the eavesdropper now observe noisy versions of the channel input. In addition, there is no secret key available as in \cite{Shannon49CommunicationTheorySecrecySystems} so that the communication must be secured solely by exploiting the properties of the noisy channel. Recently, this area of \emph{information theoretic security} or \emph{physical layer security} has drawn considerable attention especially in the area of wireless communication where it provides a promising complement to cryptographic approaches; see for example \cite{Liang09InformationTheoreticSecurity,Liu10SecuringWirelessCommunications,Bloch11InformationTheoreticSecrecy,PoorSchaefer-2017-WirelessPhysicalLayerSecurity,SchaeferBocheKhistiPoor-2017-InformationTheoreticSecurityPrivacy} and references therein. These concepts have been extended to several multi-user scenarios such as the broadcast channel \cite{CsiszarKoerner78BroadcastChannelsConfidentialMessages,Ekrem12CapacitMIMOCommonConfidential,Liu13NewResultsMIMOBCConfidential,Wyrembelski13StrongSecrecyBBC}, multiple access channel \cite{Liang08MACConfidentail,Wiese13StrongSecrecyMAC}, and interference channel \cite{Liu08InterferenceBroadcastConfidential}. All these works have in common that no secret key is available to the legitimate users.

These two approaches were combined in \cite{Yamamoto97RateDistortionShannonCipherSystem,Merhav08ShannonsSecrecySystemWithInformedReceivers,Kang10WiretapChannelSecretKey}, which study the (noisy) wiretap channel with a shared secret key. This was done from a rate-distortion point of view in \cite{Yamamoto97RateDistortionShannonCipherSystem} and \cite{Merhav08ShannonsSecrecySystemWithInformedReceivers}, while \cite{Kang10WiretapChannelSecretKey} established the secrecy capacity for the case in which no distortion is allowed at the legitimate receiver. The wiretap channel with a shared secret key and its main concepts and results are briefly reviewed in Section \ref{sec:wiretap}. Related to this problem is the wiretap channel with secured feedback, as this feedback can be used to create a shared secret key \cite{AhlswedeCai06WiretapChannelFeedback,Gunduz08SecretCommunicationFeedback,Ardestanizadeh09WiretapChannelWithFeedback}. 

A secret key shared between a transmitter and a receiver can be used to securely transmit to that receiver. If the rate of the secret key is sufficiently high, the problem becomes trivial in the single user setting (i.e., the wiretap channel with shared secret key) as the whole message can be encrypted with the secret key and the secrecy capacity is simply given by the regular channel capacity. However, when the transmitter shares independent secret keys with multiple receivers, a certain secret key can be used to securely transmit to the respective receiver, but might harm other receivers which do not share this key. Thus, multiple secret keys can result in conflicting payoffs at different receivers making it a challenging and non-trivial problem. 

Surprisingly, to the best of our knowledge the use of multiple secret keys in noisy multi-user communication scenarios has not been studied previously. Accordingly, the question of secure communication in \emph{broadcast channels (BC) with independent secret keys} is an interesting extension in this direction. In particular, this setting describes the problem of securely broadcasting a common message to two legitimate receivers, while keeping an eavesdropper ignorant of it. The transmitter shares independent secret keys of arbitrary key rates with the two receivers. This model is introduced in Section~\ref{sec:bc}.

Secure communication can now be realized by different approaches. As shared secret keys are available at the transmitter and both receivers, this suggests to follow \cite{Shannon49CommunicationTheorySecrecySystems} and use them as one-time pads to encrypt the common message. However, each receiver is aware of only one secret key. Thus, the more the secret key of one receiver is used to secure the message, the more the other receiver is hurt as the unknown secret key creates a useless message for it that interferes with the desired one. We show that when the eavesdropper has the ``strongest'' channel (i.e. both legitimate channels are degraded versions of it), the one-time pad approach is indeed capacity-achieving. More specifically, both secret keys are used to create two encrypted messages which are then accordingly encoded and transmitted with the help of a regular capacity-achieving (non-secure) broadcast channel code.

On the other hand, the properties of the noisy channels can be exploited by applying information theoretic secrecy concepts such as  \emph{wiretap coding} \cite{Liang09InformationTheoreticSecurity,Liu10SecuringWirelessCommunications,Bloch11InformationTheoreticSecrecy,PoorSchaefer-2017-WirelessPhysicalLayerSecurity,SchaeferBocheKhistiPoor-2017-InformationTheoreticSecurityPrivacy}. This approach is based on the idea of allocating some of the available rate for additional randomization to ``confuse'' the eavesdropper. The drawback of this is that it reduces the remaining rate available for the actual transmission of the confidential message. However, this approach turns out to be optimal if the eavesdropper's channel is the ``weakest'' among all channels (i.e. it is degraded with respect to both legitimate channels). To this end, the secret keys are used as \emph{fictitious messages} for the randomization. As each receiver knows one of the secret keys, the randomization overhead to be decoded at the receivers is reduced.

While in the previous cases the eavesdropper has either the strongest or the weakest channel, the much more challenging and complicated scenario occurs when one legitimate channel is stronger and one is weaker than the eavesdropper's channel. In this case, it is shown that, to achieve capacity, a careful combination of one-time pad and wiretap coding with fictitious messages is needed. The converse involves a non-trivial ``sum-rate''-like upper bound on the common message rate.

The above results are obtained in the context of discrete memoryless channels (DMCs). We further consider Gaussian channels, in which the legitimate receivers' and the eavesdropper's channel observations are impaired by additive Gaussian noise. The previous results are extended to this case resulting in a complete characterization of the capacity of the Gaussian BC with independent secret keys.

All results are stated in Section \ref{sec:results}, while the corresponding proofs are delegated to Section \ref{sec:proofs}. A conclusion is then given in Section \ref{sec:conclusion}.

\subsection*{Notation} 

Discrete random variables are denoted by capital letters and their realizations and ranges by lower case and script letters, respectively; $\N$ and $\R_+$ are the sets of positive integers and non-negative real numbers; $X-Y-Z$ denotes a Markov chain of random variables $X$, $Y$, and $Z$ in this order; for a sequence $X^n=(X_1,X_2,...,X_n)$ we frequently write $X^{i-1}=(X_1,X_2...,X_{i-1})$ and $X_{i+1}^n=(X_{i+1},X_{i+2},...,X_n)$; all logarithms, exponentials, and information quantities are with respect to the base 2; $H(\cdot)$ and $I(\cdot;\cdot)$ are the traditional entropy and mutual information; $\sP(\cdot)$ is the set of all probability distributions; $\oplus$ denotes modulo addition and $[a]^+$ abbreviates $\max\{a,0\}$; $\text{lhs}\coloneqq\text{rhs}$ assigns the value of the right hand side (rhs) of an equation to its left hand side (lhs).

\section{Wiretap Channel with Shared Secret Key}
\label{sec:wiretap}

\begin{figure}
	\centering
	\includegraphics{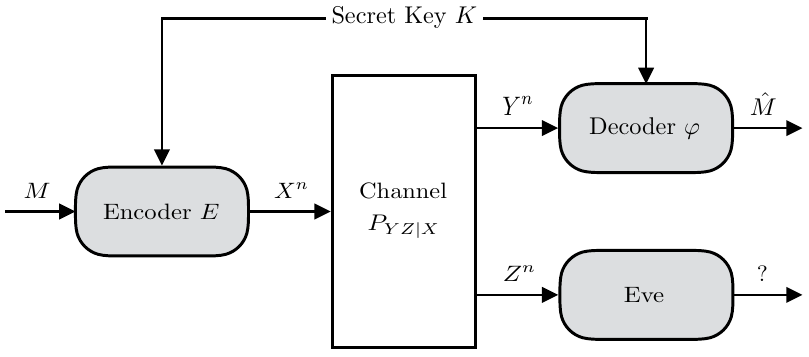}
	\caption{Wiretap channel with a shared secret key.}
	\label{fig:wiretap}
\end{figure}

In this section we review the \emph{wiretap channel with shared secret key} as depicted in Fig. \ref{fig:wiretap}. This model generalizes and combines two important communication models: Shannon's cypher system \cite{Shannon49CommunicationTheorySecrecySystems} and the classical (noisy) wiretap channel as introduced by Wyner in \cite{Wyner75WiretapChannel}. Both lead to fundamentally different approaches of how secrecy is realized. These are presented and discussed as well.

\subsection{System Model}
\label{sec:wiretap_model}

Let $\sX$, $\sY$, and $\sZ$ be finite input and output sets. For input and output sequences $x^n\in\sX$, $y^n\in\sY^n$, and $z^n\in\sZ^n$ of length $n$, the discrete memoryless wiretap channel is given by the transition probability $P_{YZ|X}^n(y^n,z^n|x^n)\coloneqq\prod_{i=1}^nP_{YZ|X}(y_i,z_i|x_i)$.

The transmitter wishes to send a message $M$ to the legitimate receiver while keeping the eavesdropper ignorant of it. The transmitter and legitimate receiver share a common secret key $K$ of rate $R_K$ which is unknown to the eavesdropper. The message $M$ and the secret key $K$ are assumed to be independent of each other and uniformly distributed over the sets $\sM\coloneqq\{1,...,M_n\}$ and $\sK\coloneqq\{1,...,K_n\}$. 

\begin{definition}
	\label{def:wiretap_code}
	An $(n,M_n,K_n)$-\emph{code} for the wiretap channel with shared secret key consists of a (stochastic) encoder
	\begin{equation}
		E: \sM\times\sK \rightarrow \sP(\sX^n)
	\end{equation}
	and a decoder at the legitimate receiver
	\begin{equation}
		\varphi: \sY^n\times\sK \rightarrow \sM.
	\end{equation}
\end{definition}

Then the average probability of decoding error at the legitimate receiver is given by
\begin{equation}
	\begin{split}
		&\bar{e}_n = \frac{1}{|\sM||\sK|}\sum_{m\in\sM}\sum_{k\in\sK}\sum_{x^n\in\sX^n} \\
		&\qquad\qquad \times\sum_{y^n:\varphi(y^n,k)\neq m}P_{Y|X}^n(y^n|x^n)E(x^n|m,k)
	\end{split}
\end{equation}
where $E(x^n|m,k)$ denotes the probability that the message $m\in\sM$ and the secret key $k\in\sK$ are encoded into the codeword $x^n\in\sX^n$.

To ensure the confidentiality of the message, we require
\begin{equation}
	I(M;Z^n) \leq \delta_n
	\label{eq:wiretap_strongsecrecy}
\end{equation}
for $\delta_n>0$ with $M$ the random variable uniformly distributed over the set of messages $\sM$ and $Z^n=(Z_1,...,Z_n)$ the channel output at the eavesdropper. This condition is termed \emph{strong secrecy} \cite{Csiszar96SecrecyCapacity,Maurer00WeakToStrongSecrecy} and the motivation is to have the total amount of information leaked to the eavesdropper small.

\begin{definition}
	\label{def:wiretap_achievable}
	For a given key rate $R_K>0$, a rate $R>0$ is an \emph{achievable secrecy rate} for the wiretap channel with shared secret key if for every $\tau>0$ there exists an $n(\tau)\in\N$ and a sequence of $(n,\Mn,K_n)$-codes such that for all $n\geq n(\tau)$ we have $\frac{1}{n}\log\Mn\geq R-\tau$, $\frac{1}{n}\log K_n=R_K$, and $I(M;Z^n)\leq\delta_n$ while $\bar{e}_n,\delta_n\rightarrow0$ as $n\rightarrow\infty$. The \emph{secrecy capacity} $C$ is given by the supremum of all achievable secrecy rates $R$.
\end{definition}

The wiretap channel with shared secret key has been studied in \cite{Yamamoto97RateDistortionShannonCipherSystem,Merhav08ShannonsSecrecySystemWithInformedReceivers,Kang10WiretapChannelSecretKey}. The secrecy capacity has been established and is restated in the following theorem.

\begin{theorem}[{\cite{Kang10WiretapChannelSecretKey}}]
	\label{the:wiretap_secrecycapacity}
	The secrecy capacity $C$ of the wiretap channel with shared secret key of rate $R_K$ is
	\begin{equation}
		C = \max_{P_{UVX}}\min\big\{[I(V;Y|U)-I(V;Z|U)]^++R_K, I(V;Y)\big\}
		\label{eq:wiretap_secrecycapacity}
	\end{equation}
	such that $U-V-X-(Y,Z)$ form a Markov chain. 
\end{theorem}

Depending on whether the term $I(V;Y|U)-I(V;Z|U)$ in \eqref{eq:wiretap_secrecycapacity} is positive or not, capacity can be achieved by using two fundamentally different approaches for realizing secrecy: the one-time pad approach or the principle of wiretap coding with fictitious messages. Both ingredients become most evident by considering either reversely degraded channels, i.e., $X-Z-Y$ form a Markov chain, or degraded channels, i.e., $X-Y-Z$ form a Markov chain. In these cases, the secrecy capacity simplifies as given in the following.

\begin{corollary}
	\label{cor:wiretap_reverselydegraded}
	The secrecy capacity $C$ of the reversely degraded wiretap channel with shared secret key of rate $R_K$ is
	\begin{equation}
	C = \max_{P_{X}}\min\big\{R_K, I(X;Y)\big\}
	\label{eq:wiretap_onetime}
	\end{equation}
	if $X-Z-Y$ form a Markov chain.
\end{corollary}

\begin{corollary}
	\label{cor:wiretap_degraded}
	The secrecy capacity $C$ of the degraded wiretap channel with shared secret key of rate $R_K$ is
	\begin{equation}
		C = \max_{P_{X}}\min\big\{I(X;Y)-I(X;Z)+R_K, I(X;Y)\big\}
	\end{equation}
	if $X-Y-Z$ form a Markov chain.
\end{corollary}

Next we discuss both approaches of one-time pad and wiretap coding in detail and show how they achieve capacity in the respective scenarios.

\subsection{One-Time Pad}
\label{sec:wiretap_onetime}

In this approach, we use the shared secret key for encryption to secure the message transmission. Following the thinking of \cite{Shannon49CommunicationTheorySecrecySystems}, the secret key can be used as a one-time pad, for which the secret key and message must be of the same rate, so that the secrecy capacity is immediately limited by the secret key rate $R_K$. Then, the encrypted message is obtained based on a modulo-$|\sK|$ addition as
\begin{equation}
	M^\oplus = M\oplus K.
	\label{eq:wiretap_xor}
\end{equation}
The encrypted message is then encoded using a classical channel encoder and transmitted over the noisy channel so that the secrecy capacity is further limited by the regular channel capacity, i.e., $\max_{P_X}I(X;Y)$. Having decoded the encrypted message $\hat{M}^\oplus$, the legitimate receiver can then use the secret key to recover the original message $\hat{M}$, i.e., $\hat{M}^\oplus\oplus K = \hat{M}\oplus K\oplus K = \hat{M}$. Since $M$ and $K$ are uniformly distributed and independent of each other, $M$ is perfectly secret from the eavesdropper, i.e., $I(M;Z^n)=0$, even if it would be able to decode the encrypted message $M^\oplus$. We conclude that the one-time pad approach achieves the rate
\begin{equation}
	R = \max_{P_X}\min\big\{R_K,I(X;Y)\big\}.
	\label{eq:wiretap_onetimerate}
\end{equation}

We observe that the rate in \eqref{eq:wiretap_onetimerate} is independent of the eavesdropper's channel and depend only on the available key rate and the legitimate channel. Accordingly, this approach works for any configuration of legitimate and eavesdropper's channels. 

In particular, in the case of reversely degraded channels, i.e., $X-Z-Y$, the one-time pad approach is optimal as it achieves the secrecy capacity as given in Corollary \ref{cor:wiretap_reverselydegraded}. However, we observe that in the case of degraded channels, i.e., $X-Y-Z$, the one-time pad approach is suboptimal as the achieved rate \eqref{eq:wiretap_onetimerate} is in general strictly smaller than the secrecy capacity as given in Corollary \ref{cor:wiretap_degraded} since $I(X;Y)-I(X;Z)>0$.

\begin{remark}
	\label{rem:noiseless}
	We immediately also obtain the corresponding result for Shannon's cypher system, which corresponds to the noiseless case, i.e., $X=Y=Z$. The secrecy capacity $C$ of the noiseless wiretap channel with shared secret key of rate $R_K$ is
	\begin{equation}
		C = \max_{P_X}\min\big\{R_K,H(X)\big\}.
		\label{eq:wiretap_onetime2}
	\end{equation}
\end{remark}

\begin{remark}
	\label{rem:onetime}
	For sufficiently large key rate, we see from \eqref{eq:wiretap_onetime} and \eqref{eq:wiretap_onetime2} that the secrecy capacity is limited by the regular channel capacity, i.e., $C=\max_{P_X}I(X;Y)$ and $C=\max_{P_X}H(X)$ respectively. The influence of the eavesdropper on the capacity completely vanishes in this case.
\end{remark}

\subsection{Wiretap Coding with Fictitious Messages}
\label{sec:wiretap_wiretap}

In contrast to the one-time pad approach, this approach exploits the noisy channel to realize secure communication. If there is no secrecy requirement, all available resources are used for message transmission. To incorporate secrecy, the crucial idea of wiretap coding \cite{Liang09InformationTheoreticSecurity,Liu10SecuringWirelessCommunications,Bloch11InformationTheoreticSecrecy,PoorSchaefer-2017-WirelessPhysicalLayerSecurity,SchaeferBocheKhistiPoor-2017-InformationTheoreticSecurityPrivacy} is now to use a certain part for randomization by introducing ``dummy'' messages. This induces multiple valid codewords for each confidential message and the one used for the actual transmission is chosen uniformly at random. The key insight is now to choose the randomization rate (i.e. the rate of the dummy message) for each confidential message according to the eavesdropper's channel quality, i.e., roughly as $I(X;Z)$. This will saturate the eavesdropper with useless information carried by the dummy messages leaving no remaining resources for decoding the confidential message itself \cite{Massey-1983-SimplifiedTreatmentWiretapChannel}. 

In the following we present a wiretap coding scheme and show how the secret key is incorporated. Let us assume that $I(X;Y)-I(X;Z)>0$ and $R_K\leq I(X;Z)$. For any input distribution $P_X\in\sP(\sX)$ we generate $|\sM||\sK||\sJ|$ independent codewords $x_{mkj}^n\in\sX^n$ where
\begin{subequations}
\begin{align}
	|\sM| &< 2^{n(I(X;Y)-I(X;Z)+R_K-2\epsilon)} \\
	|\sK| &= 2^{nR_K} \\
	|\sJ| &> 2^{n(I(X;Z)-R_K+\epsilon)}. 
\end{align}
\end{subequations}
The crucial idea here is to treat the secret key as a fictitious message for additional randomization. As the size of the secret key and the dummy randomization messages satisfy
\begin{equation}
	\frac{1}{n}\log(|\sK||\sJ|) > I(X;Z)+\epsilon,
	\label{eq:wiretap_fictitiousmessagerate}
\end{equation}
we have enough resources allocated to saturate the eavesdropper to ensure that $I(M;Z^n)\leq\delta_n$ holds, i.e., strong secrecy \eqref{eq:wiretap_strongsecrecy} is satisfied, cf. for example \cite{Bjelakovic11CapacityResultsCompoundWiretap,Bloch13SecrecyResolvability,HouKramer14EffectiveSecrecy}.

It remains to verify that the legitimate receiver can reliably decode the confidential message. Since the legitimate receiver has the secret key $k\in\sK$ as side information available, the remaining unknown indices are $m\in\sM$ and $j\in\sJ$ which are appropriately chosen to satisfy
\begin{equation}
	|\sM||\sJ| \leq 2^{n(I(X;Y)-\epsilon)}.
	\label{eq:wiretap_fictitiousreliability}
\end{equation}
It is easy to verify that this enables the legitimate receiver to reliably decode the remaining indices. 

In particular, in the case of degraded channels, i.e., $X-Y-Z$ so that $I(X;Y)-I(X;Z)>0$, wiretap coding with fictitious messages is optimal as it achieves the secrecy capacity as given in Corollary \ref{cor:wiretap_degraded}. However, we observe that in the case of reversely degraded channels, i.e., $X-Z-Y$, wiretap coding with fictitious messages is in general suboptimal (except for sufficiently large key rates for which the secrecy capacity becomes $\max_{P_{X}}I(X;Y)$).

\begin{remark}
	\label{rem:wiretap_size}
	From \eqref{eq:wiretap_fictitiousmessagerate} we observe that the maximum useful secret key rate is $I(X;Z)$. Indeed, even if $R_K>I(X;Z)$, it suffices to only use $R_K=I(X;Z)+\epsilon$ as this already saturates the eavesdropper's channel. The secrecy capacity in this case is then $\max_{P_X}I(X;Y)$ and there is no need for additional dummy messages for randomization in this case, i.e., $\sJ=\emptyset$.
\end{remark}

\begin{remark}
	\label{rem:wiretap_fictitiousreversely}
	If $I(X;Y)-I(X;Z)<0$, e.g., for reversely degraded channels $X-Z-Y$, wiretap coding with fictitious messages in general performs worse than the one-time pad approach. In particular, if $I(X;Z)\geq I(X;Y)+R_K$, no secure communication is possible at all, i.e., $R=0$, while the one-time pad approach still achieves the rate $R=\max_{P_{X}}I(X;Y)$ in this case. 
\end{remark}

From the previous reasoning and particularly Theorem \ref{the:wiretap_secrecycapacity} and Corollary \ref{cor:wiretap_degraded} we obtain the corresponding secrecy capacity for the classical wiretap channel without shared key, i.e., $R_K=0$, as stated next.

\begin{corollary}
	\label{the:wiretap_wiretap}
	The secrecy capacity $C$ of the wiretap channel is
	\begin{equation}
		C = \max_{P_{VX}}\big[I(V;Y)-I(V;Z)\big]
	\end{equation}
	such that $V-X-(Y,Z)$ form a Markov chain. If the wiretap channel is degraded, i.e., $X-Y-Z$ form a Markov chain, the secrecy capacity $C$ becomes
	\begin{equation}
		C = \max_{P_X}\big[I(X;Y)-I(X;Z)\big].
	\end{equation}
\end{corollary}

\subsection{A Combination of Both}
\label{sec:wiretap_combination}

The previous discussion has shown that neither of the two approaches of one-time pad and wiretap coding with fictitious messages is superior to the other. In fact, whenever the eavesdropper's channel is ``stronger'' than the legitimate channel ($I(X;Y)-I(X;Z)<0$) the one-time pad performs better than the wiretap coding approach and vice versa when the eavesdropper's channel is weaker ($I(X;Y)-I(X;Z)>0$).

In the following, we present a scheme that combines both approaches of one-time pad and wiretap coding. To make things non-trivial, we assume $R_K<I(X;Y)$, since otherwise the secrecy capacity is $C=\max_{P_X}I(X;Y)$ as discussed above. The main idea is now to use the secret key as a one-time pad to encrypt as much of the message as possible and use the wiretap coding approach to protect the remaining part of the message. Accordingly, we split the message $M$ into two parts
\begin{equation}
	M = (M',M^\oplus)
\end{equation}
with rates $R'$ and $R_K$, where the part $M'$ will be protected by the wiretap coding approach and $M^\oplus$ by the one-time pad approach. For any input distribution $P_X\in\sP(\sX)$ we generate $|\sM'||\sM^\oplus||\sJ|$ independent codewords $x_{m'm^\oplus j}^n\in\sX^n$ where
\begin{subequations}
	\begin{align}
		|\sM'| &< 2^{n([I(X;Y)-\max\{I(X;Z),R_K\}]^+-2\epsilon)} \\
		|\sM^\oplus| &= 2^{nR_K} \\
		|\sJ| &= \!
		\begin{cases}
			2^{n([I(X;Z)-R_K]^++\epsilon)} &\!\!\!\text{if } \!I(X;Y)\!>\!I(X;Z)\\
			0 &\!\!\!\text{if } \!I(X;Y)\!\leq\! I(X;Z).
			\end{cases}
	\end{align}
\end{subequations}
First, we observe that whenever $I(X;Y)-I(X;Z)\leq0$, the sets $\sM'$ and $\sJ$ are empty and we are falling back to the one-time pad approach as discussed in Section \ref{sec:wiretap_onetime}. If $I(X;Y)-I(X;Z)>0$, we use both approaches and secrecy is ensured similarly as in \eqref{eq:wiretap_fictitiousmessagerate} and \eqref{eq:wiretap_fictitiousreliability}. To be specific, the sets are chosen appropriately to satisfy
\begin{equation}
	\frac{1}{n}\log(|\sM^\oplus||\sJ|) > I(X;Z)+\epsilon,
\end{equation}
which gives us secrecy, and further 
\begin{equation}
	|\sM'||\sM^\oplus||\sJ| \leq 2^{n(I(X;Y)-\epsilon)},
\end{equation}
which gives us reliability.

\begin{remark}
	\label{rem:wiretap_combination}
	For both wiretap coding with fictitious messages, cf. Section \ref{sec:wiretap_wiretap}, and the combined approach, the secret key contributes to the equivocation of the eavesdropper by saturating its channel. But note that in wiretap coding with fictitious messages the secret key does not carry any useful information while in the combined approach it further carries part of the confidential message. 
\end{remark}

\begin{remark}
	\label{rem:wiretap_combination2}
	For the case $I(X;Y)-I(X;Z)>0$, both approaches have the same performance in terms of achievable secrecy rate. However, they fundamentally differ in the construction and their codebooks are of different size. In particular, while the overall codebook size of the combined approach is matched to the legitimate channel (as the receiver has to decode all three indices), wiretap coding with fictitious messages uses a codebook whose size is larger than the legitimate channel would allow, but the receiver can use its secret key as side information to reduce it to a suitable subset.
\end{remark}

\section{BC with Independent Secret Keys}
\label{sec:bc}

\begin{figure}
  \centering
  \includegraphics{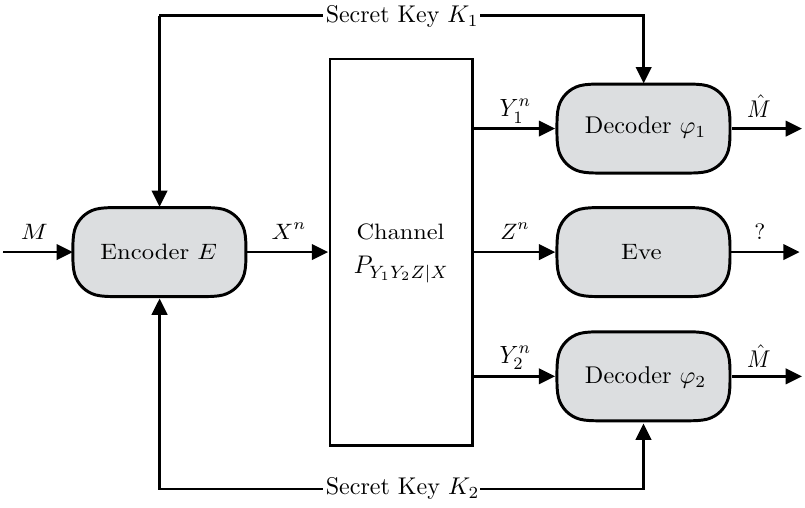}
  \caption{Broadcast channel in which the transmitter shares independent secret keys $K_1$ and $K_2$ with legitimate receivers 1 and 2 respectively. A common message $M$ is sent to both legitimate receivers keeping the eavesdropper ignorant of it.}
  \label{fig:bckeys}
\end{figure}

In the remainder of this paper we study the \emph{broadcast channel (BC) with independent secret keys} as depicted in Fig. \ref{fig:bckeys}. Let $\sX$, $\sY_1$, $\sY_2$, and $\sZ$ be finite input and output sets. For input and output sequences $x^n\in\sX^n$, $y_1^n\in\sY_1^n$, $y_2^n\in\sY_2^n$, and $z^n\in\sZ^n$ of length $n$, the discrete memoryless broadcast channel is given by the transition probability $P_{Y_1Y_2Z|X}^n(y_1^n,y_2^n,z^n|x^n)\coloneqq\prod_{i=1}^nP_{Y_1Y_2Z|X}(y_{1,i},y_{2,i},z_i|x_i)$.

Throughout the rest of the paper we study the particular communication scenario in which the transmitter broadcasts a common message $M$ to receivers 1 and 2, while keeping the eavesdropper ignorant of it. The transmitter shares independent secret keys $K_1$ and $K_2$ with receivers 1 and 2. The message and both keys are assumed to be independent of each other and uniformly distributed over the sets $\sM\coloneqq\{1,...,\Mn\}$ and $\sK_i\coloneqq\{1,...,\Kin\}$, $i=1,2$. We also write $K_{12}=(K_1,K_2)$ and $\sK_{12}=\sK_1\times\sK_2$ for short.

\begin{definition}
\label{def:bc_code}
An $(n,\Mn,\Kon,\Ktn)$-\emph{code} for the BC with independent secret keys consists of a (stochastic) encoder
\begin{equation}
	E:\sM\times\sK_1\times\sK_2\rightarrow\sP(\sX^n)
\end{equation}
and decoders at receivers 1 and 2
\begin{subequations}
\begin{align}
	\varphi_1 &: \sY_1^n\times\sK_1\rightarrow\sM, \\
	\varphi_2 &: \sY_2^n\times\sK_2\rightarrow\sM.
\end{align}
\end{subequations}
\end{definition}

Then the average probability of decoding error at receiver~$i$, $i=1,2$, is given by
\begin{equation}
\begin{split}
	&\ei = \frac{1}{|\sM||\sK_{12}|}\sum_{m\in\sM}\sum_{k_{12}\in\sK_{12}}\sum_{x^n\in\sX^n} \\
	&\qquad\quad \times\sum_{y_i^n:\varphi_i(y_i^n,k_i)\neq m}P_{Y_i|X}^n(y_i^n|x^n)E(x^n|m,k_{12}).
\end{split}
\end{equation}

As for the wiretap channel with a shared secret key, we require strong secrecy of the common message, i.e.,
\begin{equation}
	I(M;Z^n) \leq \delta_n
	\label{eq:bc_strongsecrecy}
\end{equation}
with $\delta_n>0$.

\begin{definition}
	\label{def:bc_achievable}
	A rate $R>0$ is an \emph{achievable secrecy rate} for the BC with independent secret keys if for every $\tau>0$ there exists an $n(\tau)\in\N$, key rates $R_1>0$, $R_2>0$, and a sequence of $(n,\Mn,\Kon,\Ktn)$-codes such that for all $n\geq n(\tau)$ we have $\frac{1}{n}\log\Mn\geq R-\tau$, $\frac{1}{n}\log \Kin=R_i$, $i=1,2$, and $I(M;Z^n)\leq\delta_n$ while $\eo,\et,\delta_n\rightarrow0$ as $n\rightarrow\infty$. The \emph{secrecy capacity} $C$ is given by the supremum of all achievable secrecy rates $R$.
\end{definition}

\begin{remark}
	\label{rem:keyrates}
	Note that in contrast to Definition \ref{def:wiretap_achievable}, the rates of the secret keys are not fixed and specified a priori. In particular, Definition \ref{def:bc_achievable} allows us to choose arbitrary key rates $R_1$ and $R_2$. In other words, the key rates are sufficiently high in the sense that we are provided with as much secret key as we desire.
\end{remark}

From the problem definition, in principle there are different methods possible to keep the common message secret. The shared secret keys suggest that we might use a one-time pad approach which protects the message with the help of the secret keys similarly as discussed in Section \ref{sec:wiretap_onetime}. On the other hand, the transmitter can exploit the nature of the wireless channel by using a channel code based on the idea of wiretap coding as discussed in Section \ref{sec:wiretap_wiretap}. In the following we will explore these different approaches and show that which particular approach should be used depends on the channel conditions.

\section{Main Results}
\label{sec:results}

In this section we present the main results for the BC with independent secret keys and discuss the insights that we obtain from them. The corresponding proofs are relegated to Section \ref{sec:proofs} and the appendix, respectively.

\subsection{Equal Channel Outputs}
\label{sec:results_equal}

\begin{figure*}
	\centering
	\includegraphics{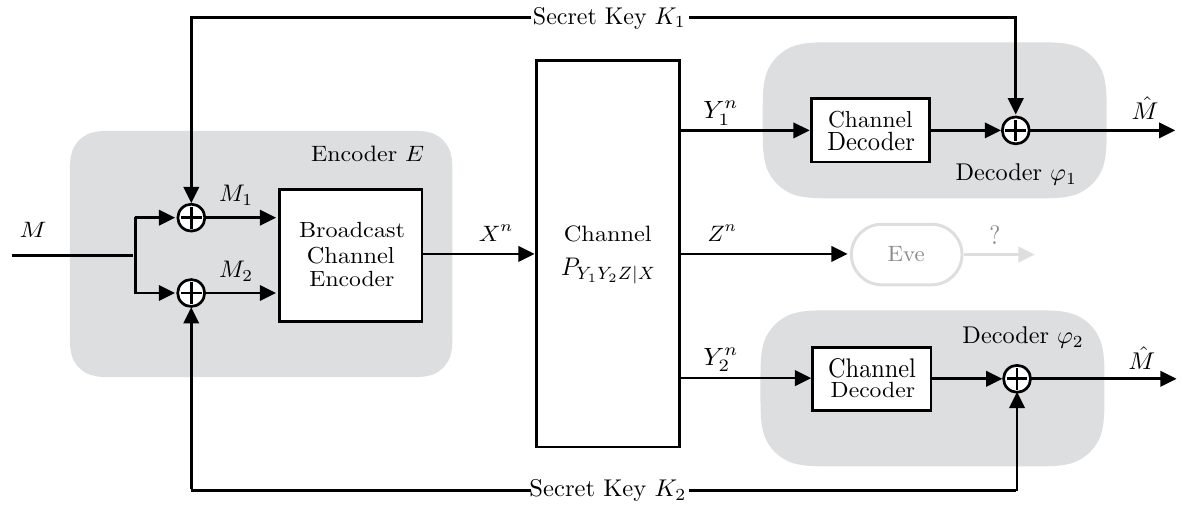}
	\caption{One-time pad approach to the BC with independent secret keys. The shared keys $K_1$ and $K_2$ are used to encrypt the common message $M$ by creating two independent messages $M_1$ and $M_2$. These two messages are encoded by a classical broadcast channel encoder and transmitted over the channel. Having decoded the individual messages $\hat{M}_1$ and $\hat{M}_2$, the legitimate receivers obtain the desired common message $\hat{M}$ with the help of the shared keys.}
	\label{fig:onetimepad}
\end{figure*}

We start with the simplest scenario, in which both legitimate receivers and the eavesdropper receive signals of the same quality and, in particular, $Y=Y_1=Y_2=Z$. 

If a legitimate receiver has no advantage over the eavesdropper at the physical layer, wiretap coding as discussed in Section \ref{sec:wiretap_wiretap} cannot be used. Accordingly, to secure the message transmission, we use the secret keys as one-time pads as presented in Section \ref{sec:wiretap_onetime} to create two \emph{encrypted} messages 
\begin{equation}
	M_1 = M \oplus K_1, \qquad\text{and}\qquad M_2 = M \oplus K_2.
	\label{eq:results_xor}
\end{equation}
These two messages are independent of each other and then encoded using a classical broadcast channel encoder and transmitted to the corresponding receivers. Having decoded the encrypted messages $M_1$ and $M_2$, each receiver can then use its own secret key to obtain the desired original message $\hat{M}$, i.e., $\hat{M}_1\oplus K_1=\hat{M} \oplus K_1\oplus K_1=\hat{M}$ and $\hat{M}_2\oplus K_2=\hat{M} \oplus K_2\oplus K_2=\hat{M}$ respectively. It becomes apparent that this one-time pad approach requires a total secret key rate of twice the common message rate. Since the message $M$ and both keys $K_1$ and $K_2$ are uniformly distributed and independent of each other, $M$ is kept perfectly secret from the eavesdropper, i.e., $I(M;Z^n)=0$, even if it would be able to decode the encrypted messages $M_1$ or $M_2$. This encoding-decoding process is depicted in Fig. \ref{fig:onetimepad}.

For the case of equal channel outputs we then obtain the following result.

\begin{theorem}
	\label{the:results_equal}
	The secrecy capacity $C$ of the BC with independent secret keys for equal channel outputs is
	\begin{equation}
		C = \max_{P_X}\frac{1}{2}I(X;Y)
		\label{eq:results_equal_capacity}
	\end{equation}
	with $Y=Y_1=Y_2=Z$, i.e., time-sharing between both legitimate receivers is optimal.
\end{theorem}
\begin{IEEEproof}
	Achievability follows from using the secret keys as one-time pads to create encrypted individual messages as in \eqref{eq:results_xor} and time-sharing between these two encrypted messages. The converse follows from Fano's inequality and the standard properties of entropy and mutual information to exploit the tension from using independent secret keys from the two receivers. A detailed proof can be found in Section~\ref{sec:proofs_equal}.
\end{IEEEproof}
\vspace*{0.5\baselineskip}

From this we immediately obtain the result for the noiseless case where the output at all receivers equals the input, i.e., $X=Y_1=Y_2=Z$. 

\begin{corollary}
	\label{cor:results_noiseless}
	The secrecy capacity $C$ of the noiseless BC with independent secret keys is
	\begin{equation}
		C = \max_{P_X}\frac{1}{2}H(X),
		\label{eq:results_equal_noiselesscapacity}
	\end{equation}
	with $X=Y_1=Y_2=Z$, i.e., time-sharing between both legitimate receivers is optimal. \qed
\end{corollary}

\begin{remark}
	\label{rem:results_timesharing}
	These results show that in the case of equal channel outputs the simple strategy of time-sharing is already capacity-achieving. Thereby, the secret keys are used as one-time pads to transform the common message into multiple individual encrypted messages.
\end{remark}

\subsection{Strongest Eavesdropper}
\label{sec:results_strong}

After having obtained some first insights for the case of equal channel outputs, we now turn to the case in which all channel outputs are of different quality but the eavesdropper has the strongest channel among all receivers in a certain sense. 

Let us first assume that the channels are reversely degraded, i.e., we have the following Markov chain: $X-Z-Y_1-Y_2$.

\begin{theorem}
	\label{the:results_reverselydegraded}
	The secrecy capacity of the BC with independent secret keys for reversely degraded channels is
	\begin{equation}
		C = \max_{P_{UX}}\min\left\{\begin{array}{l}
			I(X;Y_1|U),\\
			I(U;Y_2)
		\end{array}\right\}
		\label{eq:results_reverselydegraded_capacity}
	\end{equation}
	such that $U-X-Z-Y_1-Y_2$ form a Markov chain, i.e., superposition coding is optimal. Further, the cardinality of the range of $U$ can be bounded by $|\sU|\leq|\sX|+1$.
\end{theorem}
\begin{IEEEproof}
	Achievability follows from using the secret keys as one-time pads to create encrypted individual messages and then using superposition coding. The converse follows from Fano's inequality and standard arguments for the degraded broadcast channel. A detailed proof is given in Section~\ref{sec:proofs_reverselydegraded}.
\end{IEEEproof}

\begin{remark}
	\label{rem:superposition}
	The fact that the eavesdropper's channel is the strongest among all channels (in the sense that both legitimate channels are degraded versions of it) suggests to use the secret keys as one-time pads to secure 	the message as classical wiretap coding approaches will not work. In addition, the fact that the legitimate channels themselves can be ordered due to their degradedness suggests then to use a superposition coding scheme (as for the classical degraded BC). The previous result shows that this reasoning indeed yields a capacity-achieving coding scheme.
\end{remark}

We have seen that using the secret keys in this way essentially turns the problem of securely broadcasting a common message into the problem of broadcasting two independent individual messages. Having this observation in mind, we immediately obtain the following achievable rate which is based on the idea of Marton coding for the classical BC, cf. \cite[Sec. 8.4]{ElGamalKim-2011-NetworkInformationTheory}.

\begin{lemma}
	\label{lem:results_marton}
	An achievable secrecy rate for the BC with independent secret keys is given by
	\begin{equation}
		R = \min\left\{\begin{array}{l}
			I(U,W;Y_1), \\
			I(V,W;Y_2), \\
			\frac{1}{2}[I(U,W;Y_1) + I(V;Y_2|W) - I(U;V|W)], \\
			\frac{1}{2}[I(V,W;Y_2) + I(U;Y_1|W) - I(U;V|W)]
		\end{array}\right\}
		\label{eq:results_marton}
	\end{equation}
	for any $P_{UVWX}(u,v,w,x)$ such that $(U,V,W)-X-(Y_1,Y_2,Z)$ form a Markov chain.
\end{lemma}
\begin{IEEEproof}
	A sketch of the proof is provided in Appendix \ref{app:marton}.
\end{IEEEproof}
\vspace*{0.5\baselineskip}

We want to stress that Lemma \ref{lem:results_marton} yields an achievable secrecy rate for the BC with independent secret keys for the general case, i.e., \emph{no} constraints on the channels to the legitimate receivers and the eavesdropper are imposed. 

The previous Theorem \ref{the:results_reverselydegraded} assumed that the channels are reversely degraded. Let us now generalize this discussion in the sense that we do not impose any ordering between the legitimate channels $Y_1$ and $Y_2$. But we still assume that the eavesdropper's channel is the ``strongest'' of all channels in the sense that the following Markov chains are satisfied: $X-Z-Y_1$ and $X-Z-Y_2$. Then we obtain the following outer bound which is somewhat similar to the UV-outer bound \cite{Nair07OuterBoundBC} or the K\"orner-Marton outer bound \cite{Martion79CodingBroadcastChannel}.

\begin{theorem}
	\label{the:results_uvbound}
	The secrecy capacity $C$ of the BC with independent secret keys for reversely degraded channels $X-Z-(Y_1,Y_2)$ is upper bounded by
	\begin{align}
		C \leq \min\left\{\begin{array}{l}
			I(U;Y_1), \\
			I(V;Y_2),  \\
			\frac{1}{2}\big[I(U;Y_1)+I(X;Y_2|U)\big], \\
			\frac{1}{2}\big[I(V;Y_2)+I(X;Y_1|V)\big]\end{array}\right\}
		\label{eq:results_uvbound_bound}
	\end{align}
	for all $P_{UVX}(u,v,x)$ such that $(U,V)-X-Z-(Y_1,Y_2)$ form a Markov chain. Further, the cardinalities of the ranges of $U$ and $V$ can be bounded by $|\sU|\leq|\sX|+1$ and $|\sV|\leq|\sX|+1$.
\end{theorem}
\begin{IEEEproof}
	The proof of the upper bound in \eqref{eq:results_uvbound_bound} resembles those of the UV-bound \cite{Nair07OuterBoundBC} and K\"orner-Marton bound \cite{Martion79CodingBroadcastChannel}. A detailed proof is given in Section \ref{sec:proofs_uvbound}.
\end{IEEEproof}

\begin{remark}
	Similar to the achievability schemes discussed before, the UV-type upper bound indicates that in the case of a strongest eavesdropper, the communication problem resembles the structure of the BC with two independent individual messages (stemming from the two independent secret keys). Accordingly, similar to the classical UV-outer bound for the BC with two individual messages, our UV-type upper bound becomes tight for the previous special cases of equal channel outputs and reversely degraded channels. Thus, it recovers the converses of the corresponding Theorems \ref{the:results_equal} and~\ref{the:results_reverselydegraded}. 
\end{remark}

Finally, we want to address the case of deterministic broadcast channels, for which the channel outputs at the legitimate receivers $Y_1$ and $Y_2$  are functions of the input $X$, i.e. $Y_1=\phi_1(X)$ and $Y_2=\phi_2(X)$. For reversely degraded channels $X-Z-(Y_1,Y_2)$ we then obtain the following.

\begin{theorem}
	\label{the:results_deterministic}
	The secrecy capacity $C$ of the BC with independent secret keys for reversely degraded channels $X-Z-(Y_1,Y_2)$, for which $Y_1$ and $Y_2$ are deterministic functions of $X$, is given by
	\begin{align}
		C = \max_{P_X}\min\left\{\begin{array}{l}
			H(Y_1), \\
			H(Y_2), \\
			\frac{1}{2}H(Y_1,Y_2)
		\end{array}\right\}.
	\label{eq:results_determinstic_capacity}
	\end{align}
\end{theorem}
\begin{IEEEproof}
	Achievability follows from the Marton coding appraoch in Lemma \ref{lem:results_marton} and the converse follows from the UV-type upper bound in Theorem \ref{the:results_uvbound}. A detailed proof is given in Section~\ref{sec:proofs_deterministic}.
\end{IEEEproof}

\subsection{Weakest Eavesdropper}
\label{sec:results_weak}

\begin{figure*}
	\centering
	\includegraphics{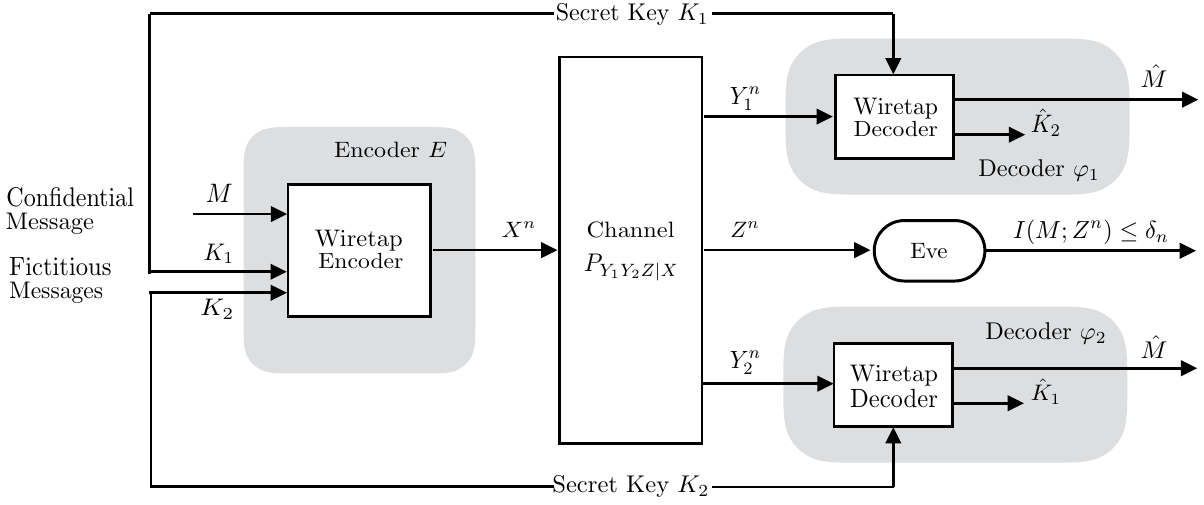}
	\caption{Wiretap coding approach to the BC with independent secret keys. The shared keys $K_1$ and $K_2$ are treated as fictitious messages and are used by the wiretap encoder as randomization resources. The wiretap decoders each have one fictitious message as side information available so that it remains for them to decode the other (useless) fictitious message and the desired common message $M$.}
	\label{fig:wiretapenc}
\end{figure*}

Next, we study the opposite case in which the eavesdropper has the weakest channel among all receivers in a certain sense. As the legitimate receivers now have an advantage over the eavesdropper at the physical layer, it is possible to use wiretap coding principles as discussed in Section \ref{sec:wiretap_wiretap}. Recall that the basic idea of wiretap coding is not to use all available rate for transmitting the message, but to allocate some of the rate to ``confuse'' the eavesdropper by applying randomized encoding strategies \cite{Liang09InformationTheoreticSecurity,Liu10SecuringWirelessCommunications,Bloch11InformationTheoreticSecrecy,PoorSchaefer-2017-WirelessPhysicalLayerSecurity,SchaeferBocheKhistiPoor-2017-InformationTheoreticSecurityPrivacy}. If a sufficient amount of rate is spent for confusion, the eavesdropper will not be able to decode the transmitted message. Obviously, the more rate is allocated to this confusion, the less rate is available for the actual transmission of the message. Here is where the shared secret keys enter the picture in this approach. They will be treated as fictitious messages playing the role of the randomization resources which are (partly) available at the legitimate receivers. This approach is visualized in Fig. \ref{fig:wiretapenc}.

In the following we consider the case in which the eavesdropper has the ``worst'' channel in the sense that its channel is degraded with respect to both legitimate channels. Thus, we assume Markov chains $X-Y_1-Z$ and $X-Y_2-Z$; however, we impose no ordering between the legitimate channels themselves.

\begin{theorem}
	\label{the:results_degraded}
	The secrecy capacity $C$ of the BC with independent secret keys and degraded channels $X-Y_1-Z$ and $X-Y_2-Z$ is
	\begin{align}
		C = \max_{P_X}\min\left\{\!\!\begin{array}{l}
			I(X;Y_1),\\
			I(X;Y_2),\\
			\frac{1}{2}[I(X;Y_1)\!+\!I(X;Y_2)\!-\!I(X;Z)]
		\end{array}\!\!\right\}\!.
	\label{eq:results_degraded_capacity}
	\end{align}
\end{theorem}
\begin{IEEEproof}
	Achievability follows by treating the secret keys as fictitious messages as part of the wiretap code. The converse follows by Fano's inequality applied to a suitably chosen sum-rate like bound. A detailed proof is given in Section \ref{sec:proofs_degraded}.
\end{IEEEproof}

\begin{remark}
	\label{rem:results_converse}
	An interesting observation is that in a capacity-achieving coding scheme, the sum rate of both secret keys must be equal to the eavesdropper's channel quality. This saturation leaves then no remaining resources for decoding of the actual common message.
\end{remark}

Next, we want to weaken the ``worst'' eavesdropper channel assumption in the sense that we study the case of less noisy channels instead of degraded channels. 

\begin{definition}
	\label{def:results_lessnoisy}
	A BC given by its joint probability $P_{YZ|X}$ is said to be \emph{less noisy} if
	\begin{equation}
		I(U;Y) \geq I(U;Z)
	\end{equation}
	holds for all $P_{UX}(u,x)$ such that $U-X-(Y,Z)$ form a Markov chain. In this case we say receiver $Y$ is less noisy than receiver $Z$ and write $Y\succeq Z$.
\end{definition}

Accordingly, we assume $Y_1\succeq Z$ and $Y_2\succeq Z$ and, again, we impose no ordering between the legitimate channels themselves. The following shows that the capacity result given in Theorem \ref{the:results_degraded} for degraded channels also holds for less noisy channels.

\begin{theorem}
	\label{the:results_lessnoisy}
	The secrecy capacity $C$ of the BC with independent secret keys and less noisy channels $Y_1\succeq Z$ and $Y_2\succeq Z$ is
	\begin{align}
		C = \max_{P_X}\min\left\{\!\!\begin{array}{l}
			I(X;Y_1),\\
			I(X;Y_2),\\
			\frac{1}{2}[I(X;Y_1)\!+\!I(X;Y_2)\!-\!I(X;Z)]
		\end{array}\!\!\right\}\!.
		\label{eq:results_lessnoisy_capacity}
	\end{align}
\end{theorem}
\begin{IEEEproof}
	Achievability follows immediately from Theorem \ref{the:results_degraded}. The converse is an adaptation of that used in Theorem~\ref{the:results_degraded} incorporating the less noisy condition. A detailed proof is given in Section \ref{sec:proofs_lessnoisy}.
\end{IEEEproof}

\subsection{Eavesdropper in the Middle}
\label{sec:results_middle}

The previous considerations either used the secret keys solely as one-time pads or solely as fictitious messages. In particular, the first approach is optimal if the eavesdropper has the ``strongest'' channel and the second approach is optimal if the eavesdropper has the ``weakest'' channel. The analysis becomes much more involved if the eavesdropper's channel is neither the strongest nor the weakest.

Accordingly, we study what happens if the eavesdropper's channel is in the middle (i.e. $X-Y_1-Z-Y_2$ form a Markov chain). That is, it is stronger than one legitimate channel, but weaker than the other one. In this case, a combination of one-time pad and fictitious messages is capacity-achieving. 

\begin{theorem}
	\label{the:results_middle}
	The secrecy capacity $C$ of the BC with independent secret keys for $X-Y_1-Z-Y_2$ is
	\begin{align}
		C = \max_{P_{UX}}\min\left\{\begin{array}{l}
		I(U;Y_2), \\
		\frac{1}{2}\big[I(X;Y_1)+I(U;Y_2)-I(U;Z)\big]\end{array}\right\}
		\label{eq:results_middle_capacity}
	\end{align}
	such that $U-X-Y_1-Z-Y_2$ form a Markov chain. The cardinality of the range of $U$ can be bounded by $|\sU|\leq|\sX|+1$.
\end{theorem}
\begin{IEEEproof}
	Achievability follows by applying a combination of the one-time pad approach and the fictitious messages approach. In particular, a superposition coding scheme is used in which the auxiliary random variable $U$ serves as a ``cloud center'' protecting the confidential message using the fictitious messages approach. On top of that, ``satellite codewords'' for the stronger receiver are superimposed protected by the one-time pad approach. The converse follows from Fano's inequality applied to a suitably chosen sum-rate like bound with a careful identification of the auxiliary random variable. A detailed proof is given in Section \ref{sec:proofs_middle}.
\end{IEEEproof}

\subsection{Gaussian Channels}
\label{sec:results_gaussian}

The results above can be extended to Gaussian channels, where input-output relations are given by 
\begin{align}
	Y_1 &= X + N_1, \\
	Y_2 &= X + N_2, \\
	Z &= X+ N_Z,
\end{align}
where the additive Gaussian noise terms $N_1$, $N_2$, and $N_Z$ have zero means and variances $\sigma_1^2$, $\sigma_2^2$, and $\sigma_Z^2$, i.e., $N_i\sim\mathcal{N}(0,\sigma_i^2)$, $i\in\{1,2,Z\}$. We consider an average transmit power constraint $P$.

Since Gaussian broadcast channels are inherently degraded, the relation among the noise variances $\sigma_1^2$, $\sigma_2^2$, and $\sigma_Z^2$ immediately determines the degradedness order. Accordingly, we extend the previous results to completely characterize the Gaussian case. W.l.o.g. we assume $\sigma_1^2\leq\sigma_2^2$, making receiver 1 the stronger of the two legitimate receivers.

\subsubsection{Strongest Eavesdropper ($\sigma_Z^2\leq\sigma_1^2\leq\sigma_2^2$)}

The eavesdropper having the smallest noise power, i.e., the ordering is $\sigma_Z^2\leq\sigma_1^2\leq\sigma_2^2$, is analogous to the Markov chain condition $X-Z-Y_1-Y_2$. Then the eavesdropper has the strongest channel as discussed in Section \ref{sec:results_strong}. In this case, the secrecy capacity is established in the following theorem.

\begin{theorem}
	\label{the:result_gaussian_strong}
	The secrecy capacity $C$ of the Gaussian BC with independent secret keys for $\sigma_Z^2\leq\sigma_1^2\leq\sigma_2^2$ is
	\begin{align}
		C = \max_{0\leq\alpha\leq1}\min 
		\left\{\begin{array}{l}
			\frac{1}{2}\log\big(1+\frac{\alpha P}{\sigma_1^2}\big) \\
			\frac{1}{2}\log\big(1+\frac{(1-\alpha) P}{\alpha P+\sigma_2^2}\big)
		\end{array}\right\}.
	\end{align}
\end{theorem}
\begin{IEEEproof}
	Achievability follows immediately from Theorem \ref{the:results_reverselydegraded} by extending the achievability scheme from the discrete memoryless case to the Gaussian case. The converse follows also from Theorem \ref{the:results_reverselydegraded} and from the converse for the classical Gaussian BC, cf. for example \cite[Theorem 5.3]{ElGamalKim-2011-NetworkInformationTheory}. The details are omitted for brevity.
\end{IEEEproof}
\vspace*{0.5\baselineskip}

\subsubsection{Weakest Eavesdropper ($\sigma_1^2\leq\sigma_2^2\leq\sigma_Z^2$)}

The eavesdropper having the highest noise power, i.e., the ordering is $\sigma_1^2\leq\sigma_2^2\leq\sigma_Z^2$, is analogous to the Markov chain condition $X-Y_1-Y_2-Z$. Then the eavesdropper has the weakest channel as discussed in Section \ref{sec:results_weak}. In this case, the secrecy capacity is established in the following theorem.

\begin{theorem}
	\label{the:result_gaussian_weak}
	The secrecy capacity $C$ of the Gaussian BC with independent secret keys for $\sigma_1^2\leq\sigma_2^2\leq\sigma_Z^2$ is
	\begin{align}
		C = \min 
		\left\{\begin{array}{l}
			\frac{1}{2}\log\big(1+\frac{P}{\sigma_2^2}\big), \\
			\frac{1}{4}\Big[\log\big(1+\frac{P}{\sigma_1^2}\big)+\log\big(1+\frac{P}{\sigma_2^2}\big)\\
			\qquad\qquad\qquad\;-\log\big(1+\frac{P}{\sigma_Z^2}\big)\Big]
		\end{array}\right\}.
	\end{align}
\end{theorem}
\begin{IEEEproof}
	Achievability follows immediately from Theorem \ref{the:results_degraded} by extending the achievability scheme from the discrete memoryless case to the Gaussian case. The converse follows also from Theorem \ref{the:results_degraded} together with the ideas for the converse of the classical Gaussian wiretap channel, cf. for example \cite[Example 22.2]{ElGamalKim-2011-NetworkInformationTheory}. The details are omitted for the sake of brevity.
\end{IEEEproof}
\vspace*{0.5\baselineskip}

\subsubsection{Eavesdropper in the Middle ($\sigma_1^2\leq\sigma_Z^2\leq\sigma_2^2$)}

When the eavesdropper's noise power satisfies $\sigma_1^2\leq\sigma_Z^2\leq\sigma_2^2$, it is analogous to the Markov chain condition $X-Y_1-Z-Y_2$. Then the eavesdropper has neither the strongest nor the weakest channel as discussed in Section \ref{sec:results_middle}. In this case, the secrecy capacity is established in the following theorem.

\begin{theorem}
	\label{the:results_gaussian_middle}
	The secrecy capacity $C$ of the Gaussian BC with independent secret keys for $\sigma_1^2\leq\sigma_Z^2\leq\sigma_2^2$ is
	\begin{align}
		C \leq \max_{0\leq\alpha\leq1}\min\left\{\!\!\!\begin{array}{l}
			\frac{1}{2}\log\big(1+\frac{\alpha P}{(1-\alpha)P+\sigma_2^2}\big), \\
			\frac{1}{4}\Big[\log\big(1+\frac{P}{\sigma_1^2}\big)+\log\big(1+\frac{\alpha P}{(1-\alpha)P+\sigma_2^2}\big)\\
			\qquad\qquad\qquad\;-\log\big(1+\frac{\alpha P}{(1-\alpha)P+\sigma_Z^2}\big)\Big]\end{array}\!\!\!\right\}\!.
		\label{eq:results_gauss}
	\end{align}
\end{theorem}
\begin{IEEEproof}
	Achievability follows immediately from Theorem \ref{the:results_middle} by extending the achievability scheme from the discrete memoryless case to the Gaussian case. The converse follows from Theorem \ref{the:results_middle} together with careful applications of the entropy power inequality and the maximum entropy theorem for Gaussian random variables. A detailed proof is given in Section \ref{sec:proofs_gaussian_middle}.
\end{IEEEproof}
\vspace*{0.5\baselineskip}

\subsubsection{Numerical Evaluation}

\begin{figure}
	\centering
	\includegraphics{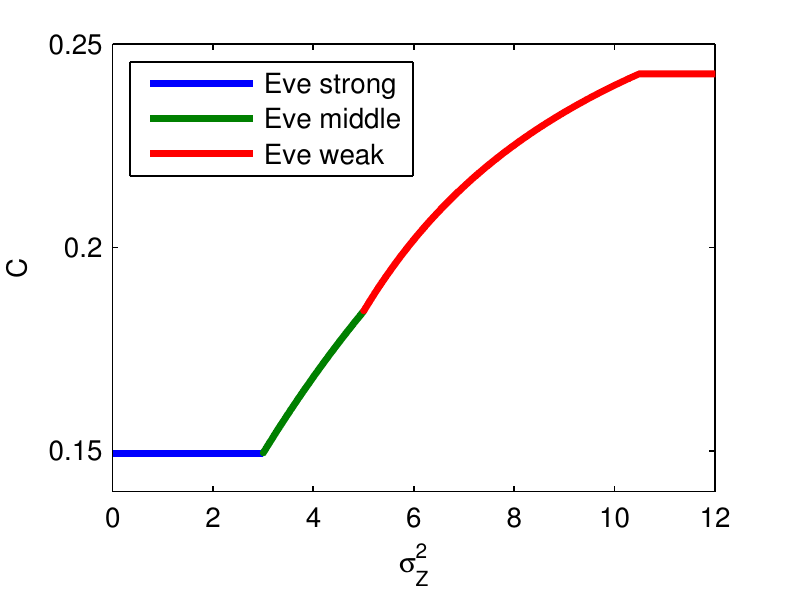}
	\caption{Secrecy capacity of the Gaussian BC with independent secret keys for $P=2$, $\sigma_1^2=3$, $\sigma_2^2=5$, and varying $\sigma_Z^2$.}
	\label{fig:sim}
\end{figure}

Here we present a numerical evaluation of the secrecy capacity of the Gaussian BC with independent secret keys. Fig. \ref{fig:sim} illustrates the behavior of the secrecy capacity for fixed noise variances of the legitimate receivers as the noise variance of the eavesdropper increases. As long as $\sigma_Z^2<\sigma_1^2$, the eavesdropper has the strongest channel and the optimal coding scheme relies solely on the one-time pad approach. Accordingly, the secrecy capacity is constant in this regime since it depends only on the fixed legitimate channels and is independent of the actual eavesdropper channel. If $\sigma_Z^2\geq\sigma_1^2$ but $\sigma_Z^2<\sigma_2^2$, the eavesdropper is in the middle and the secrecy capacity is achieved by a combination of one-time pad and wiretap coding. Finally, if $\sigma_Z^2\geq\sigma_2^2$, the eavesdropper has the weakest channel and the optimal coding scheme is wiretap coding using the secret keys as fictitious messages. At some point, the sum-rate-like bound is no longer active and the secrecy capacity is given by the capacity of the weaker link to the second receiver.

\section{Proofs}
\label{sec:proofs}

In this section we present the proofs of the main results.

\subsection{Proof of Theorem \ref{the:results_equal}}
\label{sec:proofs_equal}

As both legitimate receivers and the eavesdropper all receive channel outputs of the same quality, we use the secret keys $K_1$ and $K_2$ as one-time pads to create secure individual messages $M_1$ and $M_2$ as in \eqref{eq:results_xor}. Since $M$, $K_1$, and $K_2$ are independent of each other, we immediately have ensured that $I(M;Z^n)=0$. 
Thus, the communication problem reduces to the reliable transmission of two independent individual messages. Obviously, the rate in \eqref{eq:results_equal_capacity} is easily achievable via time-sharing. (Note that achievability also follows immediately from the Marton coding approach in Lemma \ref{lem:results_marton}. By setting $U=V=W=X$ in \eqref{eq:results_marton} we obtain the desired rate \eqref{eq:results_equal_capacity}. However, the time-sharing interpretation is more intuitive and yields valuable insights for the converse.)

Thus, it remains to show that time-sharing is already optimal. At receiver $i$, $i=1,2$, we have the following version of Fano's inequality:
\begin{equation}
	H(M|K_i,Y^n) \leq n\epsi
	\label{eq:results_equal_fano}
\end{equation}
with $\epsi\rightarrow0$ as $n\rightarrow\infty$. Making extensive use of the definition of mutual information and the chain rule, we get
\begin{align}
	nR &= H(M|K_i) \label{eq:results_equal_ind}\\
		&\leq I(M;Y^n|K_i) + n\epsi \label{eq:results_equal_fanoapp}\\
		&\leq I(M;K_i,Y^n) + n\epsi \\
		&= I(M;K_i|Y^n) + I(M;Y^n) + n\epsi \\
		&\leq I(M;K_i|Y^n) + n\epsi + n\del \label{eq:results_equal_secrecy}\\
		&= H(K_i|Y^n) - H(K_i|M,Y^n) + n\epsi + n\del \\
		&\leq H(K_i) - H(K_i|M,Y^n) + n\epsi + n\del \label{eq:results_equal_bound}
\end{align}
where \eqref{eq:results_equal_ind} follows from the independence of $M$ and $K_i$, \eqref{eq:results_equal_fanoapp} from Fano's inequality \eqref{eq:results_equal_fano}, and \eqref{eq:results_equal_secrecy} from the secrecy condition.

As \eqref{eq:results_equal_bound} must hold for both receivers simultaneously, we obtain
\begin{align}
	&nR \leq \!\min_{i\in\{1,2\}}\!\!\Big\{H(K_i) - H(K_i|M,Y^n) + n\epsi \!+ n\del\Big\} \label{eq:results_equal_minbound} \\
		&\;\leq \frac{1}{2}\big[H(K_1)+H(K_2) \nonumber \\
		&\;\qquad\quad - H(K_1|M,Y^n)-H(K_2|M,Y^n) + n\eps\big] \\
		&\;\leq \frac{1}{2}\big[H(K_{12}) - H(K_{12}|M,Y^n) + n\eps\big] \\
		&\;= \frac{1}{2}\big[H(K_{12}) + H(Y^n|M) - H(K_{12},Y^n|M) \!+\! n\eps\big] \\
		&\;= \frac{1}{2}\big[H(K_{12}) + H(Y^n|M) \nonumber \\
		&\;\qquad\quad - H(K_{12}|M) - H(Y^n|M,K_{12}) + n\eps\big] \\
		&\;= \frac{1}{2}\big[H(Y^n|M) - H(Y^n|M,K_{12}) + n\eps\big] \\
		&\;\leq \frac{1}{2}\big[H(Y^n) - H(Y^n|M,K_{12}) + n\eps\big] \\
		&\;= \frac{1}{2}\big[I(M,K_{12};Y^n) + n\eps\big] \\
		&\;\leq \frac{1}{2}\big[I(X^n;Y^n) + n\eps\big] \\
		&\;\leq \frac{1}{2}n\big[I(X;Y) + \eps\big] \label{eq:results_equal_finalbound}
\end{align}
with $\eps=\epsilon_{1,n}+\epsilon_{2,n}+2\del$ and $\eps\rightarrow0$ as $n\rightarrow\infty$. Here, the last step follows from Jensen's inequality and from defining the distribution $P_X$ in \eqref{eq:results_equal_finalbound} as the average of all $P_{X_i}$, $i=1,..,n$. This completes the converse and proves the desired result.
\qed

\subsection{Proof of Theorem \ref{the:results_reverselydegraded}}
\label{sec:proofs_reverselydegraded}

As for the equal channel case in Theorem \ref{the:results_equal} we generate individual messages $M_1$ and $M_2$ by using the available secret keys $K_1$ and $K_2$ as one-time pads, cf. \eqref{eq:results_xor}. Then the achievability of \eqref{eq:results_reverselydegraded_capacity} follows immediately by superposition coding. Here, we choose the auxiliary random variable $U$ to carry the individual message $M_1$ (as the ``cloud center'') for the weaker receiver 2. The other message $M_2$ for the stronger receiver 1 is superimposed as a ``satellite codeword'' in $X$. (Again, achievability can also be deduced from the Marton coding approach in Lemma \ref{lem:results_marton}. By setting $U=X$ and $V=\emptyset$ in \eqref{eq:results_marton} we obtain the desired rate \eqref{eq:results_reverselydegraded_capacity}.)

Again, the crucial part is the converse. We have to show that this superposition coding strategy is already optimal. Using Fano's inequality \eqref{eq:results_equal_fano} as in the proof of Theorem \ref{the:results_equal} led to \eqref{eq:results_equal_fanoapp} from which we proceed as follows:
\begin{align}
	nR &\leq I(M;Y_i^n|K_i) + n\epsi \\
		&\leq I(M,K_i;Y_i^n) + n\epsi \\
		&= I(K_i;Y_i^n|M) + I(M;Y_i^n) - I(M;Z^n) + n\eps \label{eq:results_degraded_secrecydef} \\
		&\leq I(K_i;Y_i^n|M) + n\eps \label{eq:results_degraded_secrecydeg}
\end{align}
with $\eps=\del+\epso$ where \eqref{eq:results_degraded_secrecydef} follows from the chain rule and the secrecy condition, and \eqref{eq:results_degraded_secrecydeg} from the degradedness so that $I(M;Y_i^n) - I(M;Z^n) \leq 0$, $i=1,2$.

Now, we define the auxiliary random variable
\begin{equation}
	U_i \coloneqq (M,K_2,Y_1^{i-1})
	\label{eq:results_degraded_u}
\end{equation}
and obtain for the weaker receiver 2  
\begin{align}
	nR &\leq I(K_2;Y_2^n|M) + n\eps \\
		&\leq \sum_{i=1}^nI(M,K_2;Y_{2,i}|Y_2^{i-1}) + n\eps \\
		&\leq \sum_{i=1}^nI(M,K_2,Y_2^{i-1};Y_{2,i}) + n\eps \\
		&\leq \sum_{i=1}^nI(M,K_2,Y_1^{i-1},Y_2^{i-1};Y_{2,i}) + n\eps \\
		&= \sum_{i=1}^n I(M,K_2,Y_1^{i-1};Y_{2,i}) + n\eps \label{eq:results_reverselydegraded_degraded} \\
		&= \sum_{i=1}^n I(U_i;Y_{2,i}) + n\eps 
\end{align}
where \eqref{eq:results_reverselydegraded_degraded} follows from the degradedness condition. Now, let $Q$ be a time-sharing random variable independent of all others in the model and uniformly distributed over $\{1,...,n\}$. We set $U=(U_Q,Q)$, $X=X_Q$, $Y_1=Y_{1,Q}$, and $Y_2=Y_{2,Q}$ and obtain
\begin{align}
	nR &\leq nI(U_Q;Y_{2,Q}|Q) + n\eps \\
		&\leq nI(U;Y_2)+ n\eps.
\end{align}

With the same definition of $U_i$, cf. \eqref{eq:results_degraded_u}, we obtain for the stronger receiver 1 
\begin{align}
	nR &\leq I(K_1;Y_1^n|M) + n\eps \\
		&\leq I(K_1;Y_1^n|M,K_2) + n\eps \\
		&= \sum_{i=1}^n I(K_1;Y_{1,i}|M,K_2,Y_1^{i-1}) + n\eps \\
		&\leq \sum_{i=1}^n I(K_1,X_i;Y_{1,i}|M,K_2,Y_1^{i-1}) + n\eps \label{eq:degraded_x} \\
		&= \sum_{i=1}^n I(X_i;Y_{1,i}|U_i) + n\eps \\
		&= nI(X_Q;Y_{1,Q}|U_Q,Q) + n\eps \\
		&= nI(X;Y_1|U) + n\eps
\end{align}
with $\eps=\del+\epst$ which proves the converse.

The bound $|\sU|\leq|\sX|+1$ on the cardinality of the range of the auxiliary random variable $U$ follows from the strengthened version of Carath\'eodory's theorem, cf. for example \cite{CsiszarKoerner11InformationTheory}, and standard arguments. The details are omitted for the sake of brevity. This completes the proof of the theorem.
\qed

\subsection{Proof of Theorem \ref{the:results_uvbound}}
\label{sec:proofs_uvbound}

The proof of the upper bound in \eqref{eq:results_uvbound_bound} resembles ideas of \cite{Nair07OuterBoundBC} and \cite{Martion79CodingBroadcastChannel}. Proceeding as in the proof of Theorem~\ref{the:results_reverselydegraded} we use Fano's inequality \eqref{eq:results_equal_fano} and the fact that $Y_1$ and $Y_2$ are degraded with respect to $Z$ to conclude on \eqref{eq:results_degraded_secrecydeg}, i.e., at receiver~$i$, $i=1,2$, the rate is upper bounded by
\begin{equation}
	nR \leq I(K_i;Y_i^n|M) + n\eps
	\label{eq:results_uvbound_firststep}
\end{equation}
with $\eps=\max\{\epso,\epst\}+\del$. Now, we define the auxiliary random variables
\begin{align}
	U_i &\coloneqq (M,K_1,Y_{1,i+1}^n,Y_2^{i-1}) \\
	V_i &\coloneqq (M,K_2,Y_{1,i+1}^n,Y_2^{i-1})
\end{align}
and obtain for receiver 1
\begin{align}
	nR &\leq I(K_1;Y_1^n|M) + n\eps \\
		&\leq I(M,K_1;Y_1^n) + n\eps \\
		&= \sum_{i=1}^n I(M,K_1;Y_{1,i}|Y_{1,i+1}^n) + n\eps \\
		&\leq \sum_{i=1}^n I(M,K_1,Y_{1,i+1}^n;Y_{1,i}) + n\eps \\
		&\leq \sum_{i=1}^n I(M,K_1,Y_{1,i+1}^n,Y_2^{i-1};Y_{1,i}) + n\eps \\
		&= \sum_{i=1}^n I(U_i;Y_{1,i}) +n\eps. 
\label{eq:results_uvbound_bound1}
\end{align}
Similarly, we obtain for receiver 2
\begin{align}
	nR &\leq \sum_{i=1}^n I(V_i;Y_{2,i}) +n\eps.
	\label{eq:results_uvbound_bound2}
\end{align}

We continue by considering the sum of the rates in \eqref{eq:results_uvbound_firststep} of both receivers. We obtain
\begin{align}
	&n2R \leq I(K_2;Y_2^n|M) + I(K_1;Y_1^n|M) +n\eps \\
		&\;\leq I(K_2;K_1,Y_2^n|M) + I(M,K_1;Y_1^n) +n\eps \label{eq:onetine_uvbound_question1}\\
		&\;= \sum_{i=1}^n I(K_2;Y_{2,i}|M,K_1,Y_2^{i-1}) \nonumber \\
		&\;\quad +\sum_{i=1}^n\!\Big[ I(M,K_1,Y_2^{i-1};Y_{1,i}^n) - I(M,K_1,Y_2^i;Y_{1,i+1}^n)\Big] \nonumber \\
		&\;\quad +n\eps 
\end{align}
where \eqref{eq:onetine_uvbound_question1} follows from the chain rule for mutual information and the last step follows the idea of the Gelfand-Pinsker converse in \cite{GelfandPinsker80CodingRandomParameters}, cf. also \cite[Section 7.9.2]{Kramer08MultiUserInformationTheory} and \cite[Theorem 8.5 and Appendix 8B]{ElGamalKim-2011-NetworkInformationTheory} for similar derivations. We continue as
\begin{align}
	&n2R \leq \sum_{i=1}^n \Big[I(K_2;Y_{2,i}|M,K_1,Y_2^{i-1}) \nonumber \\
		&\;\quad + I(M,K_1,Y_2^{i-1};Y_{1,i+1}^n) + I(M,K_1,Y_2^{i-1};Y_{1,i}|Y_{1,i+1}^n) \nonumber \\
		&\;\quad - I(M,K_1,Y_2^{i-1};Y_{1,i+1}^n) - I(Y_{2,i};Y_{1,i+1}^n|M,K_1,Y_2^{i-1})\Big] \nonumber \\
		&\;\quad +n\eps \\
		&\;= \sum_{i=1}^n \Big[I(K_2;Y_{2,i}|M,K_1,Y_2^{i-1}) \nonumber \\ 
		&\;\quad + I(M,K_1,Y_2^{i-1};Y_{1,i}|Y_{1,i+1}^n) \nonumber \\
		&\;\quad - I(Y_{2,i};Y_{1,i+1}^n|M,K_1,Y_2^{i-1})\Big] +n\eps \\
		&\;=\sum_{i=1}^n \Big[H(Y_{2,i}|M,K_1,Y_2^{i-1}) - H(Y_{2,i}|M,K_1,K_2,Y_2^{i-1}) \nonumber \\
		&\;\quad + H(Y_{1,i}|Y_{1,i+1}^n) - H(Y_{1,i}|M,K_1,Y_{1,i+1}^n,Y_2^{i-1}) \nonumber \\
		&\;\quad - H(Y_{2,i}|M,K_1,Y_2^{i-1}) + H(Y_{2,i}|M,K_1,Y_{1,i+1}^n,Y_2^{i-1})\Big] \nonumber \\
		&\;\quad +n\eps \\
		&\;\leq \sum_{i=1}^n \Big[H(Y_{1,i}) \!-\! H(Y_{1,i}|U_i) \!+\! H(Y_{2,i}|U_i) \!-\! H(Y_{2,i}|K_2,U_i)\Big] \nonumber \\
		&\;\quad +n\eps \\
		&\;\leq \sum_{i=1}^n \Big[I(U_i;Y_{1,i}) + H(Y_{2,i}|U_i) - H(Y_{2,i}|U_i,X_i)\Big] +n\eps \\
		&\;\leq \sum_{i=1}^n \Big[I(U_i;Y_{1,i}) + I(X_i;Y_{2,i}|U_i)\Big] +n\eps. 
\label{eq:results_uvbound_bound3}
\end{align}

Similarly, by interchanging the order of $Y_1$ and $Y_2$ we obtain
\begin{align}
	&n2R \leq I(K_1;Y_1^n|M) + I(K_2;Y_2^n|M) +n\eps \\
		&\;\leq I(K_1;K_2,Y_1^n|M) + I(M,K_2;Y_2^n) +n\eps \\
		&\;= \sum_{i=1}^n I(K_1;Y_{1,i}|M,K_2,Y_{1,i+1}^n) \nonumber \\
		&\;\quad +\sum_{i=1}^n\Big[ I(M,K_2,Y_{1,i+1}^n;Y_2^i) - I(M,K_2,Y_{1,i}^n;Y_2^{i-1})\Big] \nonumber \\
		&\;\quad +n\eps
\end{align}
where the last step follows the idea of the Gelfand-Pinsker converse, cf. \cite{GelfandPinsker80CodingRandomParameters}. We continue as
\begin{align}
	&n2R \leq \sum_{i=1}^n \Big[I(K_1;Y_{1,i}|M,K_2,Y_{1,i+1}^n) \nonumber \\
		&\;\quad + I(M,K_2,Y_{1,i+1}^n;Y_2^{i-1}) + I(M,K_2,Y_{1,i+1}^n;Y_{2,i}|Y_2^{i-1}) \nonumber \\
		&\;\quad - I(M,K_2,Y_{1,i+1}^n;Y_2^{i-1}) - I(Y_{1,i};Y_2^{i-1}|M,K_2,Y_{1,i+1}^n)\Big] \nonumber \\
		&\;\quad +n\eps \\
		&\;= \sum_{i=1}^n \Big[I(K_1;Y_{1,i}|M,K_2,Y_{1,i+1}^n) \nonumber \\
		&\;\quad + I(M,K_2,Y_{1,i+1}^n;Y_{2,i}|Y_2^{i-1})\nonumber \\
		&\;\quad  - I(Y_{1,i};Y_2^{i-1}|M,K_2,Y_{1,i+1}^n)\Big] +n\eps \\
		&\;=\sum_{i=1}^n \Big[H(Y_{1,i}|M,K_2,Y_{1,i+1}^n) - H(Y_{1,i}|M,K_1,K_2,Y_{1,i+1}^n) \nonumber \\
		&\;\quad + H(Y_{2,i}|Y_2^{i-1}) - H(Y_{2,i}|M,K_2,Y_{1,i+1}^n,Y_2^{i-1}) \nonumber \\
		&\;\quad - H(Y_{1,i}|M,K_2,Y_{1,i+1}^n) + H(Y_{1,i}|M,K_2,Y_{1,i+1}^n,Y_2^{i-1})\Big] \nonumber \\
		&\;\quad +n\eps \\
		&\;\leq \sum_{i=1}^n \Big[H(Y_{2,i}) \!-\! H(Y_{2,i}|V_i) \!+\! H(Y_{1,i}|V_i) \!-\! H(Y_{1,i}|K_1,V_i)\Big] \nonumber \\
		&\;\quad +n\eps \\
		&\;\leq \sum_{i=1}^n \Big[I(V_i;Y_{2,i}) + H(Y_{1,i}|V_i) - H(Y_{1,i}|V_i,X_i)\Big] +n\eps \\
		&\;\leq \sum_{i=1}^n \Big[I(V_i;Y_{2,i}) + I(X_i;Y_{1,i}|V_i)\Big] +n\eps. 
\label{eq:results_uvbound_bound4}
\end{align}

Now, let $Q$ be a time-sharing random variable independent of all others in the model and uniformly distributed over $\{1,...,n\}$. We set $U=(U_Q,Q)$, $V=(V_Q,Q)$, $X=X_Q$, $Y_1=Y_{1,Q}$, and $Y_2=Y_{2,Q}$ and obtain for \eqref{eq:results_uvbound_bound1}
\begin{align}
	nR &\leq \sum_{i=1}^n I(U_i;Y_{1,i}) +n\eps \\
		&= nI(U_Q;Y_{1,Q}|Q) +n\eps \\
		&\leq nI(U;Y_1) +n\eps
\end{align}
and similarly for \eqref{eq:results_uvbound_bound2}
\begin{align}
	nR &\leq nI(V;Y_2) +n\eps.
\end{align}
Accordingly, we get for \eqref{eq:results_uvbound_bound3} and \eqref{eq:results_uvbound_bound4}
\begin{align}
	&n2R \leq \sum_{i=1}^n \Big[I(U_i;Y_{1,i}) + I(X_i;Y_{2,i}|U_i)\Big] +n\eps \\
		&\quad= n\big[I(U_Q;Y_{1,Q}|Q) + I(X_Q;Y_{2,Q}|U_Q,Q)\big] +n\eps\\
		&\quad\leq n\big[I(U;Y_1) + I(X;Y_2|U)\big] +n\eps
\end{align}
and
\begin{align}
	n2R &\leq n\big[I(V;Y_2) + I(X;Y_1|V)\big] +n\eps.
\end{align}
This establishes the desired bounds in \eqref{eq:results_uvbound_bound} and it remains to bound the cardinalities as stated. This can be done exactly as for the UV-bound for the classical BC in \cite[Sec. II-C]{Nair08InnerOuterBoundsBC} and is therefore omitted for brevity. This completes the proof.
\qed

\subsection{Proof of Theorem \ref{the:results_deterministic}}
\label{sec:proofs_deterministic}

Achievability follows from the Marton coding approach in Lemma \ref{lem:results_marton}. By setting $U=Y_1$, $V=Y_2$, and $W=\emptyset$, the rate expression \eqref{eq:results_marton} simplifies to the desired rate~\eqref{eq:results_determinstic_capacity}.

The converse follows then from the UV-type upper bound in Theorem \ref{the:results_uvbound}. We observe that the first two bounds in \eqref{eq:results_uvbound_bound} can be bounded as $I(U;Y_1)\leq H(Y_1)$ and $I(V;Y_2)\leq H(Y_2)$. Further, 
\begin{align}
	&I(U;Y_1)+I(X;Y_2|U) \nonumber \\
		&\quad= I(U;Y_1) + H(Y_2|U) \label{eq:results_deterministic_conv1} \\
		&\quad= H(Y_1) - H(Y_1|U) + H(Y_2|U) \\
		&\quad= H(Y_1) + H(Y_2|Y_1) \nonumber \\
		&\quad\qquad - H(Y_2|Y_1) - H(Y_1|U) + H(Y_2|U) \\
		&\quad\leq H(Y_1,Y_2) - H(Y_2|Y_1,U) - H(Y_1|U) + H(Y_2|U) \\
		&\quad= H(Y_1,Y_2) - H(Y_1,Y_2|U) + H(Y_2|U) \\
		&\quad\leq H(Y_1,Y_2) \label{eq:results_deterministic_conv2}
\end{align}
where \eqref{eq:results_deterministic_conv1} follows from $H(Y_2|X,U)=0$ since the channel is deterministic and \eqref{eq:results_deterministic_conv2} follows from the fact that $- H(Y_1,Y_2|U) + H(Y_2|U)\leq0$. In a similar way we also get $\frac{1}{2}[I(V;Y_2)+I(X;Y_1|V)] \leq H(Y_1,Y_2)$ which give the desired bounds.
\qed

\subsection{Proof of Theorem \ref{the:results_degraded}}
\label{sec:proofs_degraded}

We start with the achievability of the rate \eqref{eq:results_degraded_capacity} and then prove the corresponding converse.

\vspace*{0.5\baselineskip}
\subsubsection{Proof of Achievability}
\label{sec:results_degraded_achievability}

The following equivalent description of \eqref{eq:results_degraded_capacity} turns out to be beneficial for the proof of achievability.

\begin{lemma}
\label{lem:results_equivalent}
The rate expression in \eqref{eq:results_degraded_capacity} can equivalently be expressed as
\begin{align}
	C = \max_{P_X}\max_{0\leq\alpha\leq1}\min\left\{\!\begin{array}{l}
		I(X;Y_1)-\alpha I(X;Z)\\
		I(X;Y_2)-(1-\alpha)I(X;Z)
	\end{array}\!\!\right\}\!.
	\label{eq:results_degraded_rewritten}
\end{align}
\end{lemma}
\begin{IEEEproof}
	A proof is provided in Appendix \ref{app:equivalent}.
\end{IEEEproof}
\vspace*{0.5\baselineskip}

Thus, instead of proving the achievability of \eqref{eq:results_degraded_capacity}, we prove the achievability of \eqref{eq:results_degraded_rewritten} for any $0\leq\alpha\leq1$. 

Next we sketch the proof of achievability. Basically, it follows the ideas of \cite{Bjelakovic13CompoundWiretap,Bloch13SecrecyResolvability,HouKramer14EffectiveSecrecy} which all present coding schemes that achieve strong secrecy as required in \eqref{eq:bc_strongsecrecy}. Accordingly, for any input distribution $P_X\in\sP(\sX)$ and $\alpha\in[0,1]$ we generate $|\sM||\sK_1||\sK_2|$ independent codewords $x_{mk_1k_2}^n\in\sX^n$ where
\begin{subequations}
\begin{align}
	|\sK_1| &> 2^{n((1-\alpha) I(X;Z)+\epsilon)} \\
	|\sK_2| &> 2^{n(\alpha I(X;Z)+\epsilon)} \\
	|\sM| &< \min\left\{\begin{array}{l}
	2^{n(I(X;Y_1)-\alpha I(X;Z)-2\epsilon)} \\
	2^{n(I(X;Y_2)-(1-\alpha) I(X;Z)-2\epsilon)}
	\end{array}\right\}.
\end{align}
\end{subequations}
The crucial idea is to use the available secret keys as randomization resources instead of generating ``dummy'' randomization indices as in the classical wiretap coding approach. As the sizes of the secret keys satisfy
\begin{equation}
	\frac{1}{n}\log(|\sK_1||\sK_2|) > I(X;Z) + 2\epsilon
\end{equation}
we have enough randomization resources to show that $I(M;Z^n)\leq\del$ holds, i.e., strong secrecy \eqref{eq:bc_strongsecrecy} is satisfied. This can be done similarly as in \cite{Bjelakovic13CompoundWiretap,Bloch13SecrecyResolvability,HouKramer14EffectiveSecrecy}.

Next, we check the reliability constraints at the legitimate receivers. Receiver 1 has the secret key $k_1\in\sK_1$ as side information available and therefore the unknown indices of the transmitted codeword are $m\in\sM$ and $k_2\in\sK_2$. As its size satisfies
\begin{equation}
	|\sM||\sK_2| \leq 2^{n(I(X;Y_1)-\epsilon)},
\end{equation}
it is straightforward to show that receiver 1 can decode the remaining indices $m\in\sM$ and $k_2\in\sK_2$. Similarly, receiver 2 has $k_2\in\sK_2$ as side information available and the unknown indices are $m\in\sM$ and $k_1\in\sK_1$ of size
\begin{equation}
	|\sM||\sK_1| \leq 2^{n(I(X;Y_2)-\epsilon)}.
\end{equation}
Again, it is easy to show that receiver 2 can decode the remaining indices $m\in\sM$ and $k_1\in\sK_1$. Thus, we conclude that \eqref{eq:results_degraded_rewritten} is an achievable rate.

\begin{remark}
	\label{rem:results_achievability}
	For classical wiretap coding, the rate needed for additional randomization is roughly $I(X;Z)$. This suffices to keep the eavesdropper ignorant of the transmitted message. The use of secret keys as the randomization resource has the advantage that parts of the needed randomization resource are already available as side information at the receivers. This reduces the loss in rate in the sense that it is only reduced by the remaining unknown randomization part (and not by the whole amount of randomization).
\end{remark}

\subsubsection{Proof of the Converse}
\label{sec:results_degraded_converse}

It remains to show the optimality of the above presented coding scheme. The first two bounds in \eqref{eq:results_degraded_capacity} are the obvious single-user bounds and follow immediately. The crucial part is to prove the third ``sum-rate''-like bound. We proceed as follows:
\begin{align}
	&n2R \leq H(M) + H(M) \label{eq:results_degraded_zero1}\\
	&\quad= H(M|K_1) + H(M|K_2) \\
	&\quad\leq I(M;Y_1^n|K_1) + I(M;Y_2^n|K_2) + n\epso\! + n\epst \label{eq:results_degraded_fano}\\
	&\quad\leq I(M;Y_1^n|K_1) + I(M;Y_2^n|K_2) \nonumber \\
	&\quad\qquad - I(M;Z^n) + n\eps \label{eq:results_degraded_secrecy}\\
	&\quad\leq I(M,K_1;Y_1^n) + I(M,K_2;Y_2^n) \nonumber \\
	&\quad\qquad - I(M;Z^n) + n\eps \\
	&\quad= I(M,K_{12};Y_1^n) + I(M,K_{12};Y_2^n) \nonumber \\
	&\quad\qquad - I(M,K_{12};Z^n) - I(K_2;Y_1^n|M,K_1) \nonumber \\
	&\quad\qquad - I(K_1;Y_2^n|M,K_2) + I(K_{12};Z^n|M) + n\eps \label{eq:results_degraded_zero2}\\
	&\quad\leq I(M,K_{12};Y_1^n) + I(M,K_{12};Y_2^n) \nonumber \\
	&\quad\qquad - I(M,K_{12};Z^n) + n\eps \label{eq:results_degraded_zero3}
\end{align}
with $\eps = \del+\epso+\epst$ and $\eps\rightarrow0$ as $n\rightarrow\infty$. Here, \eqref{eq:results_degraded_fano} follows from Fano's inequality, cf. \eqref{eq:results_equal_fano}, \eqref{eq:results_degraded_secrecy} from the secrecy criterion, and \eqref{eq:results_degraded_zero3} from the fact that $- I(K_2;Y_1^n|M,K_1) - I(K_1;Y_2^n|M,K_2) + I(K_1,K_2;Z^n|M)\leq0$. To see this last step, we write
\begin{align}
	&- I(K_2;Y_1^n|M,K_1) - I(K_1;Y_2^n|M,K_2) + I(K_{12};Z^n|M) \nonumber \\
	&\qquad = - H(K_2|M,K_1) + H(K_2|M,K_1,Y_1^n) \nonumber \\
	&\qquad\qquad - H(K_1|M,K_2) + H(K_1|M,K_2,Y_2^n) \nonumber \\
	&\qquad\qquad + H(K_{12}|M) - H(K_{12}|M,Z^n) \\
	&\qquad = H(K_2|M,K_1,Y_1^n) + H(K_1|M,K_2,Y_2^n) \nonumber \\
	&\qquad\qquad - H(K_{12}|M,Z^n) \label{eq:results_degraded_independent}\\
	&\qquad \leq H(K_2|M,K_1,Z^n) + H(K_1|M,K_2,Z^n) \nonumber \\
	&\qquad\qquad - H(K_{12}|M,Z^n) \label{eq:results_degraded_degraded}\\
	&\qquad \leq 0
\end{align}
where \eqref{eq:results_degraded_independent} follows from the fact that $M$, $K_1$, and $K_2$ are independent so that $-H(K_2|M,K_1) - H(K_1|M,K_2) + H(K_{12}|M)=0$, and \eqref{eq:results_degraded_degraded} from the Markov chain relationships $X-Y_1-Z$ and $X-Y_2-Z$ due to the degradedness. Now, with this we can proceed with the ``sum-rate'' in \eqref{eq:results_degraded_zero3} as
\begin{align}
	n2R &\leq I(M,K_{12};Y_1^n) + I(M,K_{12};Y_2^n) \nonumber \\
	&\qquad - I(M,K_{12};Z^n) + n\eps \\
	&= I(M,K_{12};Y_1^n|Z^n) + I(M,K_{12};Y_2^n) + n\eps \label{eq:results_degraded_degraded2}\\
	&\leq I(X^n;Y_1^n|Z^n) + I(X^n;Y_2^n) + n\eps \\
	&\leq n\big[I(X;Y_1|Z) + I(X;Y_2)\big] + n\eps \\
	&= n\big[I(X;Y_1) + I(X;Y_2) - I(X;Z)\big] + n\eps \label{eq:results_degraded_degraded3}
\end{align}
where \eqref{eq:results_degraded_degraded2} and \eqref{eq:results_degraded_degraded3} follow from the degradedness of the channels. This completes the proof of the converse.
\qed

\subsection{Proof of Theorem \ref{the:results_lessnoisy}}
\label{sec:proofs_lessnoisy}

The proof of achievability follows immediately from Theorem \ref{the:results_degraded} so that it remains to prove the converse for the less noisy case. In the following we will highlight the differences from the converse proof for the degraded case in Theorem \ref{the:results_degraded}.

Following the lines \eqref{eq:results_degraded_zero1}-\eqref{eq:results_degraded_zero3} for the ``sum-rate'' term we also obtain
\begin{align}
	n2R &\leq I(M,K_{12};Y_1^n) + I(M,K_{12};Y_2^n) \nonumber \\
		&\quad - I(M,K_{12};Z^n) - I(K_2;Y_1^n|M,K_1) \nonumber \\
		&\quad - I(K_1;Y_2^n|M,K_2) + I(K_{12};Z^n|M) + n\eps \\
		&\leq I(M,K_{12};Y_1^n) + I(M,K_{12};Y_2^n) \nonumber \\
		&\quad - I(M,K_{12};Z^n) + n\eps \label{eq:results_lessnoisy_lessnoisy1}
\end{align}
with $\eps = \del+\epso+\epst$ and $\eps\rightarrow0$ as $n\rightarrow\infty$. Here, \eqref{eq:results_lessnoisy_lessnoisy1} follows again from the fact that $- I(K_2;Y_1^n|M,K_1) - I(K_1;Y_2^n|M,K_2) + I(K_1,K_2;Z^n|M)\leq0$. To see that this also holds for less noisy channels, we proceed as follows. With $I(K_1,K_2;Z^n|M)=I(K_1;Z^n|M)+I(K_2;Z^n|M,K_1)$ we observe that
\begin{equation}
	-I(K_2;Y_1^n|M,K_1)+I(K_2;Z^n|M,K_1) \leq 0
	\label{eq:results_lessnoisy_lemma1}
\end{equation}
which is a consequence of the less noisy condition $Y_1\succeq Z$, cf. for example \cite[Proposition 1]{MansourSchaeferBoche14SecrecyBCRSIJointIndividual}. With $I(K_1;Z^n|M)\leq I(K_1;Z^n|M,K_2)$ we also have
\begin{equation}
	-I(K_1;Y_2^n|M,K_2)+I(K_1;Z^n|M,K_2) \leq 0
	\label{eq:results_lessnoisy_lemma2}
\end{equation}
which validates the correctness of the inequality in \eqref{eq:results_lessnoisy_lessnoisy1}.

Now, we can proceed with the ``sum-rate'' in \eqref{eq:results_lessnoisy_lessnoisy1} as
\begin{align}
	&n2R \leq I(M,K_{12};Y_1^n) + I(M,K_{12};Y_2^n) \nonumber \\
	&\;\qquad - I(M,K_{12};Z^n) + n\eps \\
	&\;\leq \sum_{i=1}^n\Big[I(M,K_{12};Y_{1,i}|Y_1^{i-1}) - I(M,K_{12};Z_i|Z_{i+1}^n)\Big] \nonumber \\
	&\;\qquad + I(X^n;Y_2^n) + n\eps \\
	&\;=\sum_{i=1}^n\Big[I(M,K_{12};Y_{1,i}|Y_1^{i-1},Z_{i+1}^n) \nonumber \\
	&\;\qquad -\! I(M,K_{12};Z_i|Y_1^{i-1},Z_{i+1}^n)\Big] \!+\! I(X^n;Y_2^n) \!+\! n\eps \label{eq:results_lessnoisy_csisarsum} \\
	&\;=\sum_{i=1}^n\Big[I(V_i;Y_{1,i}|U_i) - I(V_i:Z_i|U_i)\Big] \!+\! I(X^n;Y_2^n) \!+\! n\eps \label{eq:results_lessnoisy_aux}
\end{align}
where \eqref{eq:results_lessnoisy_csisarsum} follows from the Csisz\'ar sum identity \cite[Lemma~7]{CsiszarKoerner78BroadcastChannelsConfidentialMessages} and \eqref{eq:results_lessnoisy_aux} from the definition of auxiliary random variables $U_i\coloneqq(Y_1^{i-1},Z_{i+1}^n)$ and $V_i\coloneqq (M,K_{12},U_i)$. Now, let $Q$ be a time-sharing random variable independent of all others in the model and uniformly distributed over $\{1,...,n\}$. We set $U=(U_Q,Q)$, $V=V_Q$, $X=X_Q$, $Y_1=Y_{1,Q}$, and $Z=Z_Q$ to obtain
\begin{align}
	n2R	&\leq n\big[I(V;Y_1|U)-I(V;Z|U) + I(X;Y_2)\big] \!+\! n\eps  \label{eq:results_lessnoisy_singleletter}\\
		&\leq n\big[I(V^*;Y_1)-I(V^*;Z) + I(X;Y_2)\big] \!+\! n\eps \label{eq:results_lessnoisy_v}\\
		&= n\big[I(X;Y_1) - I(X;Y_1|V^*) \nonumber \\
		&\qquad - I(X;Z) + I(X;Z|V^*) + I(X;Y_2)\big] \!+\! n\eps \\
		&\leq n\big[I(X;Y_1) + I(X;Y_2) - I(X;Z)\big] \!+\! n\eps \label{eq:results_lessnoisy_lessnoisy}
\end{align}
where \eqref{eq:results_lessnoisy_v} follows from letting $V^*$ be a random variable distributed according to the choice $U=u^*$ that maximizes the difference in \eqref{eq:results_lessnoisy_singleletter}. Finally, the inequality in \eqref{eq:results_lessnoisy_lessnoisy} follows from the less noisy condition $Y\succeq Z$ which implies $I(X;Y_1|V^*)\geq I(X;Z|V^*)$. This concludes the proof of the converse.
\qed

\subsection{Proof of Theorem \ref{the:results_middle}}
\label{sec:proofs_middle}

\subsubsection*{Proof of Achievability}

For  achievability we want to exploit both previously discussed approaches: secret keys as one-time pads and secret keys as fictitious messages in wiretap codes. Therefore, we split the message $M$ into two parts
\begin{equation}
	M=(M',M^\oplus)
\end{equation}
with rates $R'$ and $R^\oplus$, where the part $M'$ will be protected by the wiretap coding approach and $M^\oplus$ by the one-time pad approach. Accordingly, we split the secret keys as
\begin{equation}
	K_1=(K_1',K_1^\oplus) \quad\text{and}\quad K_2=(K_2',K_2^\oplus)
\end{equation}
such that $R^\oplus=R_{K_1}^\oplus=R_{K_2}^\oplus$ to ensure that $M^\oplus$ can be protected following the one-time pad idea by creating encrypted messages
\begin{equation}
	M_1^\oplus=M^\oplus\oplus K_1^\oplus \quad\text{and}\quad M_2^\oplus=M^\oplus\oplus K_2^\oplus
\end{equation}
both of rate $R^\oplus$.

We use a superposition coding scheme with two layers. For the random code construction, for any $P_U$ we first generate ``cloud centers'' carrying $M'$, $K_1'$, $K_2'$, and $M_2^\oplus$. Here, $M'$ will be protected by a wiretap code approach. Then, according to $P_{X|U}$ we generate ``satellite codewords'' carrying $M_1^\oplus$. 

Now, the stronger receiver 1 decodes the cloud center (having $K_1'$ as side information available) and the satellite codeword, while the weaker receiver 2 decodes only the cloud center (having $K_2'$ as side information available). If we choose the rates of all messages such that they satisfy
\begin{align}
	R'+R_{K_2}'+R^\oplus &\leq I(U;Y_1) \label{eq:results_middle_ach1}\\
	R'+R_{K_1}'+R^\oplus &\leq I(U;Y_2) \\
	R^\oplus &\leq I(X;Y_1|U) \\
	R_{K_1}'+R_{K_2}'+R^\oplus &> I(U;Z) \\
	R^\oplus &> I(X;Z|U) \label{eq:results_middle_ach4}
\end{align}
the average probabilities of error become arbitrarily small. Having decoded $M_1^\oplus$ and $M_2^\oplus$ respectively, the receivers obtain the desired part $M^\oplus$ with the help of their side information $K_1^\oplus$ and $K_2^\oplus$ respectively. 

It remains to check the security of the transmitted message. As $R_{K_1}'+R_{K_2}'+R^\oplus > I(U;Z)$ and $R^\oplus > I(X;Z|U)$, the rate of the fictitious messages is high enough to show that $I(M';Z^n)\leq\del$ holds, i.e., strong secrecy \eqref{eq:bc_strongsecrecy} is satisfied. This can be done similarly as in \cite{Bjelakovic13CompoundWiretap,Bloch13SecrecyResolvability,HouKramer14EffectiveSecrecy}. Note that no information about $M^\oplus$ is leaked to the eavesdropper due to the one-time pad $K_2^\oplus$ even if the eavesdropper is able to decode the encrypted message $M_2^\oplus$. The second part $M_1^\oplus$ in the satellite codeword is immediately protected by the one-time pad $K_1^\oplus$. 

Now, by applying Fourier-Motzkin elimination to get rid of $R'$, $R^\oplus$, $R_{K_1}'$, and $R_{K_2}'$ in \eqref{eq:results_middle_ach1}-\eqref{eq:results_middle_ach4} having in mind that $R=R'+R^\oplus$, yields
\begin{align}
	R \leq \min\left\{\begin{array}{l}
		I(U;Y_1), \\
		I(U;Y_2), \\
		\frac{1}{2}\big[I(X;Y_1)+I(U;Y_2)-I(U;Z)\big]
	\end{array}\right\}
\end{align}
where we observe that the first condition can be omitted as $I(U;Y_2)\leq I(U;Y_1)$ holds due to the degradedness. This concludes the proof of achievability.

\begin{remark}
	\label{rem:middleachievability}
	We want to highlight the difference from the previous approach in Theorem \ref{the:results_degraded}. There, the available secret keys are solely used as fictitious messages. From a legitimate receiver's perspective, using its own secret key as a fictitious message increases the equivocation rate at the eavesdropper but does not contribute to the message rate. However, Theorem~\ref{the:results_degraded} shows that for degraded channels $X-Y_1-Z$ and $X-Y_2-Z$ this is already sufficient to achieve capacity, cf. Theorem \ref{the:results_degraded}.
	
	The one-time pad approach enters the picture when one legitimate receiver is stronger than the eavesdropper. Additionally to using the secret keys as fictitious messages, one part of the secret key is used to create an encrypted message $M_2^\oplus$ which is also used as a fictitious message part to increase the equivocation rate. Again, it does not contribute to the message rate of receiver 1, but it is useful for receiver 2. To resolve the mismatch, the stronger receiver 1 receives the missing part encrypted in the satellite codeword.
	
	This approach is in particular beneficial if receiver 2 is much weaker than the eavesdropper while receiver 1 and the eavesdropper are close in terms of channel quality.
\end{remark}

\subsubsection*{Proof of the Converse}
We define the auxiliary random variable
\begin{equation}
	U_i \coloneqq (M,K_2,Z^{i-1}).
	\label{eq:results_middle_qu}
\end{equation}
The first bound in \eqref{eq:results_middle_capacity} is a single-user bound and follows as
\begin{align}
	nR &\leq I(M;Y_2^n|K_2) + n\epso \label{eq:results_middle_fano} \\
		&\leq I(M,K_2;Y_2^n) + n\epso \label{eq:results_middle_y2} \\
		&\leq \sum_{i=1}^n I(M,K_2;Y_{2,i}|Y_2^{i-1}) + n\epso \\
		&\leq \sum_{i=1}^n I(M,K_2,Y_2^{i-1};Y_{2,i}) + n\epso \\
		&\leq \sum_{i=1}^n I(M,K_2,Y_2^{i-1},Z^{i-1};Y_{2,i}) + n\epso \\
		&\leq \sum_{i=1}^n I(M,K_2,Z^{i-1};Y_{2,i}) + n\epso \\
		&= \sum_{i=1}^n I(U_i;Y_{2,i}) + n\epso \label{eq:results_middel_aux}
\end{align}
with $\epso\rightarrow0$ as $n\rightarrow\infty$. Here, \eqref{eq:results_middle_fano} follows from Fano's inequality and \eqref{eq:results_middel_aux} from the definition of $U_i$, cf. \eqref{eq:results_middle_qu}.

The crucial part is the second ``sum-rate''-like bound. From the proof of Theorem \ref{the:results_degraded} in \cite{Schaefer14CommonMessageSecretKeys}, we know that $n2R \leq I(M,K_1;Y_1^n) + I(M,K_2;Y_2^n)  - I(M;Z^n) + n\eps$. Following this, we obtain
\begin{align}
	n2R &\leq I(M,K_1;Y_1^n) + I(M,K_2;Y_2^n)  \nonumber \\
		&\qquad- I(M;Z^n) + n\eps \\
		&= I(M,K_{12};Y_1^n) + I(M,K_2;Y_2^n) - I(M,K_2;Z^n) \nonumber \\
		&\qquad - I(K_2;Y_1^n|M,K_1) + I(K_2;Z^n|M) + n\eps \\
		&\leq I(M,K_{12};Y_1^n) + I(M,K_2;Y_2^n) \nonumber \\
		&\qquad- I(M,K_2;Z^n) + n\eps
	\label{eq:results_middle_sumbound}
\end{align}
with $\eps=\del+\epso+\epst$ and $\eps\rightarrow0$ as $n\rightarrow\infty$. Here, \eqref{eq:results_middle_sumbound} follows from the fact that $- I(K_2;Y_1^n|M,K_1) + I(K_2;Z^n|M)\leq0$. Now we proceed as
\begin{align}
	n2R &\leq I(M,K_{12};Y_1^n) + I(M,K_2;Y_2^n) \nonumber \\
		&\qquad - I(M,K_2;Z^n) + n\eps \\
		&= I(M,K_{12};Y_1^n) + I(M,K_2;Y_2^n) - I(M,K_{12};Z^n)\nonumber \\
		&\qquad + I(K_1;Z^n|M,K_2) + n\eps 
\end{align}
where the first and third term are bounded as
\begin{align}
	&I(M,K_{12};Y_1^n) - I(M,K_{12};Z^n) \nonumber \\
		&\qquad = I(M,K_{12};Y_1^n|Z^n) \leq I(X^n;Y_1^n|Z^n) \label{eq:results_middle_degraded}\\
		&\qquad \leq n\big[I(X;Y_1|Z)\big] = n\big[I(X;Y_1) - I(X;Z)\big].
\end{align}
Here, \eqref{eq:results_middle_degraded} holds due $X-Y_1-Z$. The second term becomes
\begin{align}
	&I(M,K_2;Y_2^n) \leq  \sum_{i=1}^n I(U_i;Y_{2,i})
\end{align}
which follows as in \eqref{eq:results_middle_y2}-\eqref{eq:results_middel_aux}. And finally, the last term is
\begin{align}
	&I(K_1;Z^n|M,K_2) = \sum_{i=1}^n I(K_1;Z_i|M,K_2,Z^{i-1}) \\
		&\quad\leq \sum_{i=1}^nI(K_1,X_i;Z_i|M,K_2,Z^{i-1}) \\
		&\quad= \sum_{i=1}^nI(X_i;Z_i|M,K_2,Z^{i-1}) = \sum_{i=1}^nI(X_i;Z_i|U_i).
\end{align}
Now, by following standard arguments of introducing a time-sharing random variable independent of all others and uniformly distributed over $\{1,...,n\}$, it is straightforward to finally obtain $R\leq I(U;Y_2) + \epso$ and $2R\leq \big[I(X;Y_1) + I(U;Y_2) - I(U;Z)\big] + \eps$.

The bound $|\sU|\leq|\sX|+1$ on the cardinality of the range of the auxiliary random variable $U$ follows from the strengthened version of Carath\'eodory's theorem, cf. \cite{CsiszarKoerner11InformationTheory}. \qed

\subsection{Proof of Theorem \ref{the:results_gaussian_middle}}
\label{sec:proofs_gaussian_middle}

Achievability follows from the corresponding scheme for discrete memoryless channels in Theorem \ref{the:results_middle}, where we choose $U\sim\alpha P$ (wiretap coding), $V\sim(1-\alpha)P$ (additional one-time pad part), and $X=U+V$ with $0\leq\alpha\leq1$ parametrizing the power allocation between the two schemes. 

The converse follows from an application of the entropy power inequality and the maximum entropy theorem. We write the first condition in \eqref{eq:results_middle_capacity} as
\begin{equation}
	I(U;Y_2) = h(Y_2) - h(Y_2|U)
\end{equation}
with $h(\cdot)$ the differential entropy. By the maximum entropy theorem, the first term is upper bounded by $h(Y_2)\leq\frac{1}{2}\log(2\pi e(P+\sigma_2^2))$. For the second term $h(Y_2|U)$ we observe
\begin{align}
	\frac{1}{2}\log(2\pi e\sigma_2^2) &= h(N_2) = h(Y_2|X) \leq h(Y_2|U) \\
	&\leq h(Y_2) \leq \frac{1}{2}\log(2\pi e(P+\sigma_2^2))
\end{align}
so that there must exist an $\alpha\in[0,1]$ such that
\begin{equation}
	h(Y_2|U) = \frac{1}{2}\log\big(2\pi e((1-\alpha)P+\sigma_2^2)\big).
	\label{eq:results_gauss_alpha}
\end{equation}
With this, we obtain for the first expression in \eqref{eq:results_middle_capacity} 
\begin{align}
	I(U;Y_2) &\leq \frac{1}{2}\log\Big(1+\frac{\alpha P}{(1-\alpha)P+\sigma_2^2}\Big)
	\label{eq:results_uy2}
\end{align}
which already establishes the first bound in \eqref{eq:results_gauss} as desired.

With $I(X;Z)=I(U;Z)+I(X;Z|U)$ we can write the second condition in \eqref{eq:results_middle_capacity} as
\begin{align}
	I(X;Y_1) + I(U;Y_2) - I(X;Z) + I(X;Z|U).
\end{align}
In the following, we will bound each term separately. From \eqref{eq:results_uy2} we already know that $I(U;Y_2) \leq \frac{1}{2}\log(1+\frac{\alpha P}{(1-\alpha)P+\sigma_2^2})$. Following the proof for the classical Gaussian wiretap channel by using the entropy power inequality, we further get
\begin{equation}
	I(X;Y_1) - I(X;Z) \leq \frac{1}{2}\log\Big(1+\frac{P}{\sigma_1^2}\Big) - \frac{1}{2}\log\Big(1+\frac{P}{\sigma_Z^2}\Big).
\end{equation}
Thus, it remains to bound the last term $I(X;Z|U)$ as
\begin{align}
	I(X;Z|U) &= h(Z|U) - h(Z|U,X) \\
		&= h(Z|U) - \frac{1}{2}\log(2\pi e\sigma_Z^2).
\end{align}
To bound the remaining term, we apply the entropy power inequality as follows:
\begin{align}
	h(Y_2|U) &= h(Z+\tilde{N}|U) \\
		&\geq \frac{1}{2}\log\Big(2^{2h(Z|U)}+2^{2h(\tilde{N}|U)}\Big) \\
		&= \frac{1}{2}\log\Big(2^{2h(Z|U)} + 2\pi e(\sigma_2^2-\sigma_Z^2)\Big)
\end{align}
where $\tilde{N}$ is Gaussian distributed with variance $\sigma_2^2-\sigma_Z^2$. Now, we choose $\alpha$ as in \eqref{eq:results_gauss_alpha} to obtain
\begin{equation}
	2\pi e((1-\alpha)P+\sigma_2^2) \geq 2^{2h(Z|U)}+2\pi e(\sigma_2^2-\sigma_Z^2)
\end{equation}
so that
\begin{align}
	h(Z|U) &\leq \frac{1}{2}\log\big(2\pi e((1-\alpha)P+\sigma_Z^2)\big).
\end{align}
Now, putting everything together yields
\begin{align}
	&I(X;Y_1)+I(U;Y_2)-I(U;Z) \nonumber \\
		&\qquad \leq \frac{1}{2}\Big[\log\Big(1+\frac{P}{\sigma_1^2}\Big)+\log\Big(1+\frac{\alpha P}{(1-\alpha)P+\sigma_2^2}\Big)\nonumber \\
		&\qquad\qquad\qquad-\log\Big(1+\frac{\alpha P}{(1-\alpha)P+\sigma_Z^2}\Big)\Big]
\end{align}
which yields the remaining bound in \eqref{eq:results_gauss}. This completes the converse and therewith the proof.
\qed

\section{Conclusion}
\label{sec:conclusion}

A secret key shared between a transmitter and a receiver clearly helps to securely communicate in the presence of an eavesdropper. In the ``single-user'' user scenario of the wiretap channel with shared secret key, the gain in secrecy capacity can immediately be quantified. It corresponds to the key rate of the shared secret key which agrees with common intuition. The same question becomes much more involved and non-trivial to answer for more complex communication scenarios in which multiple secret keys are shared among the legitimate users. Shared secret keys suggest the use of one-time pads to encrypt confidential messages for keeping external eavesdroppers ignorant. However, a secret key shared between the transmitter and one receiver might harm other receivers which do not share this key. Thus, multiple secret keys can result in conflicting payoffs at different receivers, which raises the question of how these keys should be used in an optimal way. In this paper, we have studied the BC with independent secret keys in which a common message must be securely transmitted to two legitimate receivers each of which shares an independent secret key with the transmitter. This is the simplest extension beyond the single-user wiretap channel that captures the above mentioned effects.

For situations in which the eavesdropper's channel is ``stronger'' than the legitimate channels, classical wiretap coding does not work and the message must be protected by using the secret keys as one-time pads. For several special cases such as the noiseless BC, BC with equal channel outputs, or degraded BC, it has been shown that this strategy is actually optimal and the secrecy capacity has been established. On the other hand, in situations in which the eavesdropper's channel is the ``weakest'' channel, it has been shown that it is optimal to use the secret keys not as one-time pads but as fictitious messages for randomization in wiretap coding. The corresponding secrecy capacity has been established as well. The most challenging case is when the eavesdropper has neither the strongest nor the weakest channel. Here, a careful combination of the one-time pad approach and wiretap coding with fictitious messages is needed to establish the secrecy capacity for this case. First, these problems have been addressed for the situation in which all channels are discrete memoryless. Then, Gaussian channels have been considered and the previous results have been extended to this case as well yielding a complete characterization of the secrecy capacity of the Gaussian BC with independent secret keys. 

For future work, it would be interesting to study how these results obtained for the case of two legitimate receivers extend to an arbitrary number of legitimate receivers. Another direction for future work would be to characterize how large of secret key rates are needed to achieve capacity and what happens if the secret key rates are bounded below those rates.

\appendix

\subsection{Sketch of Proof of Lemma \ref{lem:results_marton}}
\label{app:marton}

As described in Section \ref{sec:results_equal}, the secret keys are used as one-time pads to create individual encrypted messages $M_1=M\oplus K_1$ and $M_2=M\oplus K_2$, cf. \eqref{eq:results_xor} and Fig. \ref{fig:onetimepad}, which are then encoded and transmitted. As $M$, $K_1$, and $K_2$ are independent of each other, we immediately have $I(M;Z^n)=0$ even if the eavesdropper would able to decode the encrypted messages $M_1$ or $M_2$. 

Now, the achievability of the rate given in \eqref{eq:results_marton} follows from the idea of Marton coding for the classical BC, cf. \cite[Sec. 8.4]{ElGamalKim-2011-NetworkInformationTheory}. In particular, the created individual messages $M_1$ and $M_2$ are encoded following this approach which yields the bounds on the rate as given in \eqref{eq:results_marton}, cf. \cite[Proposition 8.1]{ElGamalKim-2011-NetworkInformationTheory}.
\qed

\subsection{Proof of Lemma \ref{lem:results_equivalent}}
\label{app:equivalent}

With the function
\begin{align}
	f(t) \!=\! \begin{cases}
		(1\!-\!t)I(X;Y_1) + t[I(X;Y_2)-I(X;Z)]	&\text{if } t\leq \frac{1}{2} \\
		(1\!-\!t)[I(X;Y_1)-I(X;Z)] + tI(X;Y_2)	&\text{if } t\geq \frac{1}{2}
	\end{cases}
\end{align}
we can express the rate expression in \eqref{eq:results_degraded_capacity} as 
\begin{align}
	C = \max_{P_X}\min_{0\leq t\leq1}f(t).
	\label{eq:results_degraded_duality}
\end{align}
As the function $f(t)$ is piecewise linear, it is sufficient to evaluate it at the corner points, i.e., when $t=0$, $t=1$, and $t=\frac{1}{2}$, to convince ourself that the rate expressions in \eqref{eq:results_degraded_capacity} and \eqref{eq:results_degraded_duality} are equivalent.

Now we have to show that \eqref{eq:results_degraded_duality} is equivalent to the desired expression \eqref{eq:results_degraded_rewritten}. Therefore, we rewrite \eqref{eq:results_degraded_rewritten} (where we omit the outer maximization for short) as
\begin{align}
	&\max_{0\leq\alpha\leq1}\min_{0\leq t\leq1} \Big[(1-t)\big[I(X;Y_1)-\alpha I(X;Z)\big] \nonumber \\
	&\qquad\qquad\qquad +t\big[I(X;Y_2)-(1-\alpha)I(X;Z)\big]\Big] \nonumber \\
	\begin{split}
		=&\min_{0\leq t\leq1}\max_{0\leq\alpha\leq1} \Big[(1-t)\big[I(X;Y_1)-\alpha I(X;Z)\big]  \\
		&\qquad\qquad\qquad +t\big[I(X;Y_2)-(1-\alpha)I(X;Z)\big]\Big] \label{eq:results_degraded_minmax}
	\end{split}
\end{align}
where the equality follows from the minimax theorem. Now eliminating $\alpha$ in \eqref{eq:results_degraded_minmax} yields for $t\leq\frac{1}{2}$ and $t\geq\frac{1}{2}$ the corresponding expressions in \eqref{eq:results_degraded_duality} which are then equivalent to the original formulation \eqref{eq:results_degraded_capacity}.
\qed


\begin{IEEEbiography}
	[{\includegraphics[width=1in,height=1.25in,clip,keepaspectratio]{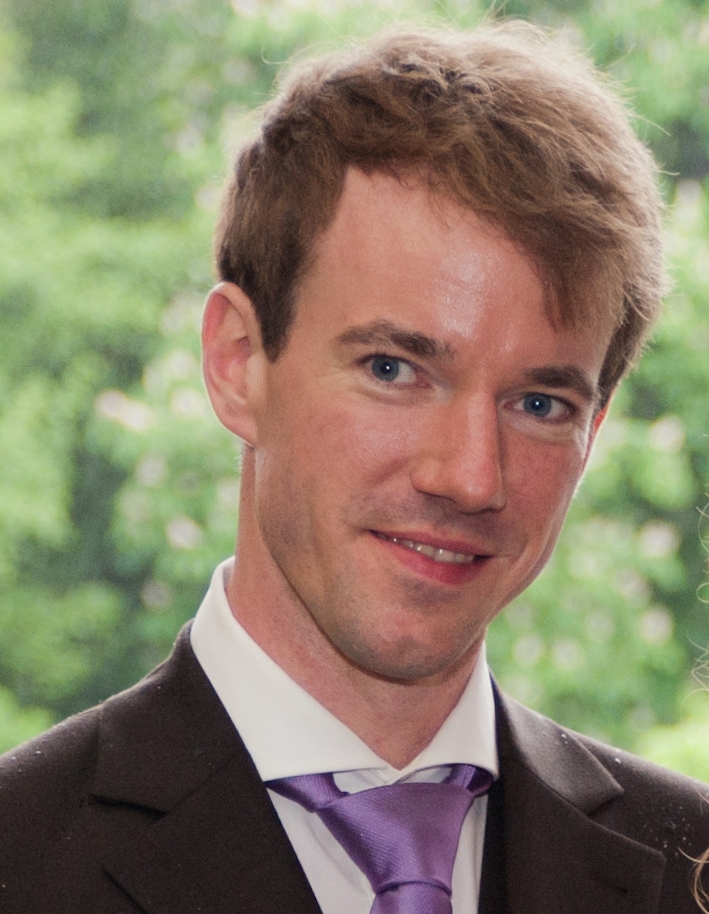}}]{Rafael F. Schaefer}
	(S'08--M'12--SM'17) received the Dipl.Ing. degree in electrical engineering and computer science from the Technische Universit\"at Berlin, Germany, in 2007, and the Dr.-Ing. degree in electrical engineering from the Technische Universit\"at M\"ünchen, Germany, in 2012. From 2007 to 2010, he was a Research and Teaching Assistant with the Technische Universit\"at Berlin and from 2010 to 2013, with the Technische Universit\"at M\"ünchen. From 2013 to 2015, he was a Post-Doctoral Research Fellow with Princeton University. Since 2015, he has been an Assistant Professor with the Technische Universit\"at Berlin. Among his publications is the recent book \emph{Information Theoretic Security and Privacy of Information Systems} (Cambridge University Press, 2017). He was a recipient of the VDE Johann-Philipp-Reis Prize in 2013. He received the best paper award of the German Information Technology Society in 2016. He was one of the exemplary reviewers of the \textsc{IEEE Communication Letters} in 2013. He is an Associate Member of the IEEE Information Forensics and Security Technical Committee. He is currently an Associate Editor of the \textsc{IEEE Transactions of Communications}.
\end{IEEEbiography}

\begin{IEEEbiography}
	[{\includegraphics[width=1in,height=1.25in,clip,keepaspectratio]{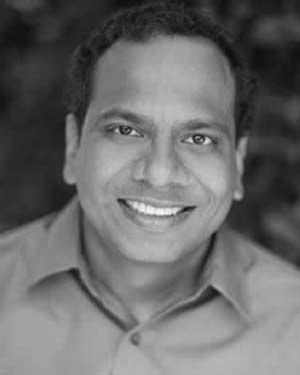}}]{Ashish Khisti}
	(S'01--M'09) received the B.A.Sc. degree from the Engineering Sciences Program (Electrical Engineering Option) the University of Toronto, Toronto, ON, Canada, in 2002, and the M.S. and Ph.D. degrees in electrical engineering and computer science from the Massachusetts Institute of Technology, Cambridge, MA, USA, in 2004 and 2009, respectively. He is currently an Associate Professor with the University of Toronto and holds a Canada Research Chair in wireless networks. His research interests include information theory, physical layer security, and error control coding for multimedia applications. He also actively consults telecommunication companies. He was a recipient of the HP-IRP Award from Hewlett-Packard and an Ontario Early Researcher Award. He currently serves as an Associate Editor in Shannon theory for the \textsc{IEEE Transactions on Information Theory}.
\end{IEEEbiography}
\balance
\begin{IEEEbiography}
	[{\includegraphics[width=1in,height=1.25in,clip,keepaspectratio]{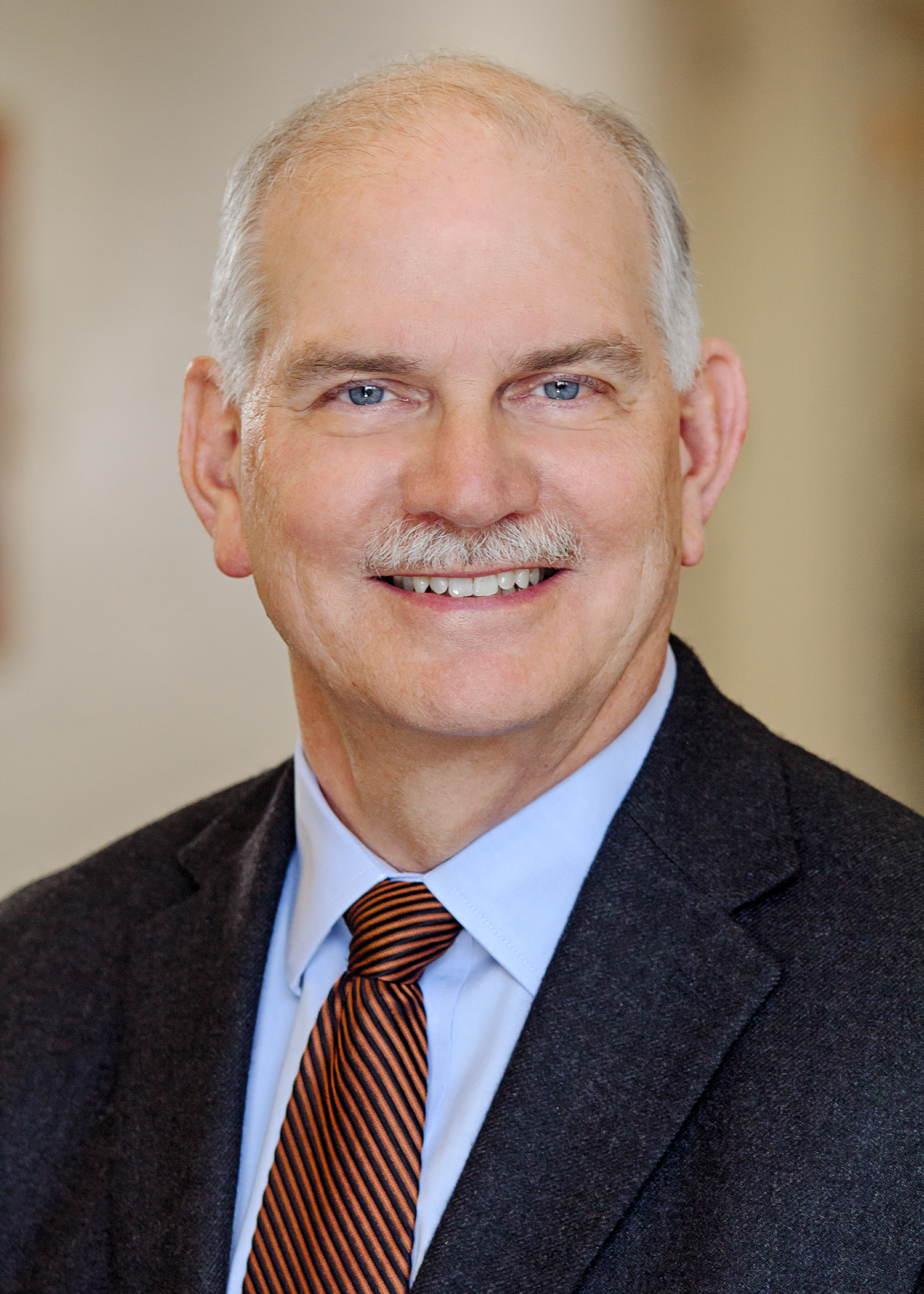}}]{H. Vincent Poor}
	(S'72--M'77--SM'82--F'87) received the Ph.D. degree in EECS from Princeton University in 1977. From 1977 to 1990, he was on the faculty of the University of Illinois at Urbana–Champaign. Since 1990, he has been on the faculty at Princeton, where he is currently the Michael Henry Strater University Professor of Electrical Engineering. From 2006 to 2016, he served as the Dean of the Princeton’s School of Engineering and Applied Science. He has also held visiting appointments at several other universities, including most recently at Berkeley and Cambridge. His research interests are in the areas of information theory and signal processing, and their applications in wireless networks and related fields, such as energy systems and social networks. Among his publications in these areas is the recent book \emph{Information Theoretic Security and Privacy of Information Systems} (Cambridge University Press, 2017).

	Dr. Poor is a Member of the National Academy of Engineering and the National Academy of Sciences and a Foreign Member of the Royal Society. He is also a Fellow of the American Academy of Arts and Sciences, the National Academy of Inventors, and other national and international academies. He received the Marconi Award and the Armstrong Award of the IEEE Communications Society in 2007 and 2009, respectively. Recent recognition of his work includes the 2017 IEEE Alexander Graham Bell Medal, Honorary Professorships at Peking University and Tsinghua University, both conferred in 2017, and the D.Sc. \emph{honoris causa} from Syracuse University in 2017.
\end{IEEEbiography}

\end{document}